# Time-average properties of $z \sim 0.6$ major mergers: mergers significantly scatter high-$z$ scaling relations

M. Puech [1]★ H. Flores,[1] M. Rodrigues,[2] F. Hammer[1] and Y. B. Yang[1]

[1]*GEPI, Observatoire de Paris, PSL University, CNRS, 5 Place Jules Janssen, F-92190 Meudon, France*
[2]*Astrophysics, Department of Physics, Denys Wilkinson Building, Keble Road, Oxford OX1 3RH, UK*

**ABSTRACT**

Interpreting the scaling relations measured by recent large kinematic surveys of $z \lesssim 1$ galaxies has remained hampered by large observational scatter. We show that the observed ISM and morpho-dynamical properties along the average $z \sim 0.6$ major merger describe a very self-consistent picture in which star formation is enhanced during first passage and fusion as a result of gravitational perturbations due to the interaction, while the gas velocity dispersion is simultaneously enhanced through shocks that convert kinematic energy associated with bulk orbital motions into turbulence at small scales. Angular momentum and rotation support in the disc are partly lost during the most perturbing phases, resulting in a morphologically compact phase. The fractions of present-day E/S0 versus later type galaxies can be predicted within only a few per cent, confirming that roughly half of local discs were reformed in the past 8–9 Gyr after gas-rich major mergers. Major mergers are shown to strongly scatter scaling relations involving kinematic quantities (e.g. the Tully–Fisher or Fall relations). Selecting high-$z$ discs relying only on $V/\sigma$ turns out to be less efficient than selecting discs from multiple criteria based on their morpho-kinematic properties, which can reduce the scatter of high-$z$ scaling relations down to the values measured in local galaxy samples.

## 1 INTRODUCTION

Spatially resolved observations of distant massive galaxies (hereafter defined as $M_{\rm stellar} \geq 10^{10}\,{\rm M}_\odot$) have largely impacted our view of galaxy evolution (see e.g. review of first-generation IFU surveys by Glazebrook 2013; see also Wisnioski et al. 2015, Stott et al. 2016, Rodrigues et al. 2017 to cite only a few of more recent results). These surveys have revealed a wealth of information on the evolution of the morpho-kinematic properties of distant galaxies and the evolution of their scaling relations (see summary and references in Section 2).

Amongst these samples, the IMAGES-CDFS sample of star-forming galaxies at $z \sim 0.6$ is of particular interest for studying the progenitors of local spiral galaxies; since most of the star formation at $z \leq 1$ occurred in such galaxies (Bell et al. 2005; Hammer et al. 2005), and because the fraction of E/S0 did not rise between $z \sim 0.6$ and $z = 0$ (Delgado-Serrano et al. 2010; Puech et al. 2012, hereafter P12), $z \leq 1$ intermediate-mass galaxies are the likeliest progenitors of local spirals (Hammer et al. 2005). While the IMAGES-CDFS sample remains of limited size with only 33 galaxies, it was shown to be unbiased relative to the global population of intermediate-mass galaxies at this epoch. The sample was selected as $I_{AB} \leq 23.5$ and $EW_0({\rm [O\,II]}) \geq 15$ Å to guarantee their detection by the multi-IFU spectrograph FLAMES/GIRAFFE at VLT (Yang et al. 2008), which resulted in a sample that is fully representative of the $J$-band luminosity function at $z = 0.5\text{–}1.0$ (Yang et al. 2008). At these redshifts, the near-infrared luminosity functions of the global population and of the blue subpopulation of galaxies are very similar (Cirasuolo et al. 2007), which implies that at first order, emission-line and blue galaxies sample the same population (Delgado-Serrano et al. 2010). This guarantees that the IMAGES-CDFS sample is representative of the star-forming population of galaxies at these epochs and can be used to infer any property of the global underlying population in the same range of stellar mass (i.e. $M_{\rm stellar} = 10^{10-11}\,{\rm M}_\odot$). As expected, this sample was indeed shown to be representative of the intermediate-mass population at $z \sim 0.6$ in terms of SFR/stellar mass density, *SFR* versus $M_{\rm stellar}$ (the so-called main sequence of star- forming galaxies, Puech et al. 2014, hereafter P14), and rotation velocities $V_{\rm rot}$ versus $M_{\rm stellar}$ [the Tully–Fisher relation (TFR), Puech et al. 2010]. This paper further illustrates that this sample follows other classical scaling relations (including their scatter) such as the $j_{\rm disc}$ versus $M_{\rm stellar}$ (the Fall relation), size versus $M_{\rm stellar}$, or between the gas velocity dispersion $\sigma_{\rm gas}$ and *SFR*.

★ E-mail: mathieu.puech@obspm.fr

The morpho-kinematic (spatially resolved) properties of the IMAGES-CDFS galaxies were used to show that both observations and models predict that ~50 per cent of local spiral galaxies have undergone a major merger since $z \sim 1.5$ (P12). The same sample was used to study the evolution of *SFR* as a function of time along an average merging sequence at $z \sim 0.6$ and illustrate that the scatter of the *SFR*–$M_{stellar}$ relation does not preclude that (gas-rich) major mergers played a major role in the most recent structural evolution of local spiral galaxies (P14). Here we generalize this study to the evolution of other ISM and dynamical properties as a function of time along the same average merging sequence.

While the impact of mergers is more and more recognized as an important and perhaps dominant driver for the formation of present-day local spiral galaxies (see Section 2), their impact on several important scaling relations, which could carry potentially important information on different key processes associated with galaxy evolution and formation, may have been underweighted. For instance, deviations from the expected evolution in the discs specific angular momentum $j_{disc}$ (versus mass or proxies) could trace the impact of different processes affecting the gas in distant galaxies (e.g. in/outflows and/or mergers, see Puech et al. 2007; Romanowsky & Fall 2012; Fall & Romanowsky 2013; Obreschkow & Glazebrook 2014; Burkert et al. 2016). Variations during major mergers occurring at high $z$ can result in a scatter in $j_{disc} \sim 0.3$ dex (Puech et al. 2007), which could contribute significantly to the observed scatter in the $j_{disc}$ versus $M_{stellar}$ relation at high $z$ (~0.2–0.3 dex, e.g. Harrison et al. 2017; Swinbank et al. 2017). A second example is provided by the evolution of the gas velocity dispersion $\sigma_{gas}$ with redshift, for which several models were proposed to explain how turbulence can be sustained to such a high level in high-$z$ galaxies (e.g. Krumholz & Burkhart 2016). The large scatter in $\sigma_{gas}$ measurements was once again shown to hamper the comparison between the different scaling relations predicted by these models (Johnson et al. 2018). Another example is provided by the scatter of the TFR that dramatically increases with $z$, which complicates the measurement of the evolution in zero-point and comparison with models (e.g. Tiley et al. 2016; Turner et al. 2017; Übler et al. 2017). This scatter in velocity was shown to be associated with the enlarged fraction of galaxies with perturbed morphologies and/or kinematics (including mergers) at $z \sim 0.6$ and above (Flores et al. 2006; Kassin et al. 2007; Covington et al. 2010; Puech et al. 2010).

Better understanding of how fundamental properties evolve *on average* along a typical major merger could provide important insights into the enlarged scatter observed in many important scaling relations between galaxy properties at high $z$. We investigate in this paper how the most important properties characterizing the ISM and the overall morphology and dynamics evolve along a typical average major merger at $z \sim 0.6$. Since the last major merger was shown to be an important factor is setting the final morphology of a large fraction of present-day local spiral galaxies (e.g. Hopkins et al. 2010c; Puech et al. 2012; Font et al. 2017), this will also provide a useful benchmark for models to check how realistic simulated major mergers are at $z < 1$, and what is the quantitative impact of disc regrowth in local discs.

This paper is organized as follows: in Section 2, we review the role of major mergers in the evolution of disc galaxies at $z \leq 1$, which should interest the reader that is seeking a complete census of the impact of these processes on the population of disc galaxies at relatively recent epochs. In Section 3, we describe the sample, how the different physical properties studied in this paper were measured, and how their time-average variation along a typical $z \sim 0.6$ major merger were constructed. In Section 4, we describe the resulting time-average trends and how they compare with different results (from both observations and simulations) in the literature. In Section 5, we discuss their co-evolution and show that the predicted bulge-to-total ratio (*B/T*) distribution for the merger remnants, which can be predicted from the $z \sim 0.6$ major merger properties, provides a very good match to observed fractions of early-type and late-type galaxies in the local universe. We also discuss the significant impact that the temporal variation of several physical quantities have on several scaling relations widely used in the literature. Throughout the paper, a standard concordance cosmological model was assumed, while a diet Salpeter initial mass function (IMF) was adopted (Bell et al. 2003).[1]

## 2 BACKGROUND: REVIEWING THE ROLE OF MAJOR MERGERS IN THE EVOLUTION OF DISC GALAXIES AT $z < 1$

### 2.1 Major mergers as evidence of the predicted hierarchical assembly of galaxies

Mergers between dark matter haloes are an inherent process of the hierarchical Lambda cold dark matter ($\Lambda$-CDM) model (White & Rees 1978). Observational evidence was gathered over the past two decades by measuring the rate at which galaxies inhabiting haloes merge as a function of time, using different techniques. Although most often limited to massive galaxies and to major coalescences (stellar mass ratios $\mu_{stellar} \geq 1/4$), larger and/or deeper surveys are now allowing such measurements to be conducted out to $z \sim 3$–4 (Mundy et al. 2017; Mantha et al. 2018), while the deepest studies can now probe the low-mass end of the luminosity function including at very high $z$ (Ventou et al. 2017).

The different techniques used to measure the merger rate (i.e. pair counts, morphological disturbances or more recently from kinematic peculiarities, see e.g. P12, López-Sanjuan et al. 2013; Rodrigues et al. 2018) are prone to different (systematic) uncertainties, which can limit their accuracy by up to a factor ~10 (Hopkins et al. 2010d; Rodriguez-Gomez et al. 2016). Combining different methods is a key ingredient for getting accurate measurements since it allows breaking the degeneracies that a single technique faces for identifying galaxies in all the different merger phases (i.e. pre-fusion/pairs of galaxies, fusion/merging systems, post-fusion/virialization remnants; see e.g. Patton et al. 2000; Hung et al. 2014, 2015; Pawlik et al. 2016; Snyder et al. 2017; see also Hammer, Flores & Rodrigues 2017). The measured and theoretically predicted (major) merger rates are then found to agree remarkably well, which provides strong support to the hierarchical assembly of galaxies in the $\Lambda$-CDM model, at least out to $z < 1.5$ (P12; Rodrigues et al. 2018).

Major mergers between massive galaxies are therefore a strongly observationally supported prediction of the hierarchical $\Lambda$-CDM model but there are numerous processes that can impact the formation and evolution of galaxies (Somerville & Davé 2015). Although the major merger rate is found to evolve significantly with redshift (as $\sim(1+z)^{1-3}$ at least up to $z \sim 1.5$,[2] e.g. López-Sanjuan et al. 2015), deconvolving the impact of mergers on galaxy formation and evolution from other processes might actually be easier at low redshifts, since at least gas accretion from intergalactic filaments is predicted to be much reduced at $z < 1$ (Kereš et al. 2009; Nelson et al. 2013). While the major merger rate was significantly reduced in the past Gyr compared to the early Universe, both observations and models predict that ∼50–60 per cent (depending on mass) of local spiral galaxies have undergone a major merger since $z \sim 1.5$ (P12; Rodrigues et al. 2018). This implies that major mergers cannot be neglected to study the most recent evolution of present-day disc galaxies at $z \lesssim 1$.

---

[1] The difference ($-0.093$ dex, Gallazzi et al. 2008) with the widely used Chabrier (2003) IMF for quantities such as the star formation rate or stellar mass is small and does not impact this study.

[2] A possible flattening of the major merger rate at $z \sim 2$ was indeed reported (Ryan et al. 2008; Man, Zirm & Toft 2016; Mundy et al. 2017).

### 2.2 Major mergers as a driver of the structural evolution of local massive galaxies

Numerical simulations have shown for decades how major mergers (involving progenitors with low gas contents) can result in remnants with properties very similar to elliptical galaxies (e.g. Toomre & Toomre 1972; Negroponte & White 1983; Barnes 1988; Hernquist 1992; Naab, Burkert & Hernquist 1999; Bournaud, Jog & Combes 2005; Burkert et al. 2008; Bois et al. 2011; Naab et al. 2014). More recently, simulations also suggested that *gas-rich* major mergers can rebuild large rotating discs as observed in local spiral galaxies (Springel & Hernquist 2005; Robertson et al. 2006; Hopkins et al. 2009b; Stewart et al. 2009; Richard et al. 2010; Athanassoula et al. 2016; Sparre & Springel 2016). Explicit examples of disc reformation following gas-rich mergers have been identified in cosmological simulations (Governato et al. 2009; Brook et al. 2012b; Aumer et al. 2013; Aumer, White & Naab 2014; Kannan et al. 2015; Rodriguez-Gomez et al. 2016; Font et al. 2017; Sparre & Springel 2017; Martin et al. 2018), and supported by observations (Hammer et al. 2005, 2009a; Peirani et al. 2009; Puech et al. 2009).

Over the past years, several possible tensions between the measured major merger rate and the expected role of these mergers on the evolution of galaxy structures were resolved. Since it was thought for long that major mergers systematically destroy the progenitor discs during the process, the large fraction of spiral galaxies amongst the local (massive) population (∼70 per cent, e.g. Delgado-Serrano et al. 2010) was suggested to be in conflict with the measured and predicted major merger rate (see above), which was dubbed as the 'disc survival issue' (Stewart et al. 2008). However, since gas-rich mergers allow large discs to reform after fusion, there is actually no need for discs to survive in Λ-CDM; the morpho-kinematic properties of $z \sim 0.6$ galaxies indeed suggest that 50 per cent of massive local galaxies have rebuilt their discs during the past 8–9 Gyr (P12, see also discussion in Section 4.2).

Another suggested issue was related to the internal structure of the central bulges. In the local universe, 80 per cent of galaxies more massive than $M_{\text{stellar}} = 10^{10}$ M$_\odot$ show a significant bulge in their central part (i.e. with bulge-to-total flux ratios B/T > 0.05, Fisher & Drory 2011; see also Kormendy et al. 2010). While 'classical' bulges are considered to be the result of galaxy mergers (as a result of the violent relaxation of stars in the central part of the remnant, Lynden-Bell 1967), 'pseudo-bulges' are rather believed to be the product of secular evolution corresponding to smoother dynamical processes in which the gas is brought much more gradually to the centre (Kormendy & Kennicutt 2004; Athanassoula 2005; Debattista et al. 2006; Heller, Shlosman & Athanassoula 2007). The fact that pseudo-bulged (or bulgeless) galaxies are present in at least half of large nearby spiral galaxies (Weinzirl et al. 2009; Kormendy et al. 2010) was therefore suggested to be in conflict with the measured merger rate.

Sauvaget et al. (2018) conducted a systematic study of gas-rich major merger remnants at $z \sim 0$ using hydrodynamical simulations that take into account realistic initial conditions for the progenitors at $z \sim 1.5$. In particular, they adopted initial gas fractions consistent with observational constraints (e.g. Rodrigues et al. 2012), which results in gas fractions larger than 50 per cent in the progenitors. Major mergers between these progenitors result mostly in pseudo-bulges and only a few classical ones. Such mergers do not prevent secular processes to occur in the remnants; the gas left after fusion gradually falls into the centre forming bars and pseudo-bulges for most of the orbital parameters after a few Gyr. The formation of bulge-dominated galaxies of the Milky Way mass requires events with more stringent conditions, i.e. extremely small gas fractions and preferentially 1:1 mergers (Sauvaget et al. 2018).

In summary, gas-rich major mergers occurring in the past 9 Gyr of about half of local spiral galaxies have probably resulted in the reformation of their discs. This is consistent with a general inside-out growth of discs in local spirals as deduced from the study of their stellar populations (e.g. Neichel et al. 2008; González Delgado et al. 2015).

### 2.3 The role of major mergers in the evolution of the SFR density

The exact role of major mergers as a possible driver of the evolution of the star formation rate density has been widely discussed. The relatively small scatter (∼0.3 dex) of the *SFR*–$M_{\text{stellar}}$ relation (often dubbed as the 'main sequence' of star-forming galaxies, hereafter MS) at least up to $z \sim 2.5$ (e.g. Noeske et al. 2007; Rodighiero et al. 2011; Whitaker et al. 2012) suggested that SFR enhancements in simulated major mergers were too large in amplitude and too short in duration to account for the MS scatter, preventing mergers to drive a significant part of the *SFR* density. However, when comparing the MS scatter with a statistical representative sample of star-forming galaxies at $z \sim 0.6$, the resulting average *SFR* peak during the fusion phase is found to be consistent with the average SFR enhancement duration (∼1.8 Gyr) and amplitude (enhancement of a factor ∼2–3) as inferred from the MS scatter (P14). Recent hydrodynamical simulations including more and more realistic feedback recipes have now produced *SFR* enhancement peaks during the fusion phase that are consistent with MS scatter (e.g. Hopkins et al. 2013). Moreover, it has been realized that major mergers do not uniquely correspond to strong starbursts scattered above the MS but that they can be also associated with more moderate star-forming galaxies spread across the MS (P14; Sparre & Springel 2017; Cibinel et al. 2018).

It was also claimed that major mergers cannot be an important driver of the evolution of the *SFR* density because the enhancement of the star formation activity directly measured in morphologically selected major mergers was estimated to be typically less than 10 per cent of the *SFR* density (e.g. Robaina et al. 2009; Rodighiero et al. 2011; Lamastra et al. 2013; Martin et al. 2017). The fraction of *SFR* associated with a particular merger phase is found to be broadly consistent with a large range of results at $z < 1$ from the literature (P14);[3] galaxies selected in the fusion phase are found to be associated with $23 \pm 5$ per cent of the total *SFR*, while morphologically selected mergers (i.e. mostly galaxies in the fusion peak; Hopkins et al. 2010b; Lotz et al. 2010) are consistent with 15–21 per cent of the *SFR* at similar redshifts and masses (see

---

[3] Including the morphological demographics of Luminous InfraRed Galaxies (hereafter, LIRGs) at $z < 1$.

Bell et al. 2005; Jogee et al. 2009; Robaina et al. 2009). When corrected from possible contributions of other processes (such as internal instabilities, minor mergers, or gas accretion), the *SFR* peak *enhancement* due to major mergers is found to be $10 \pm 5$ per cent, which is consistent with results reported by Robaina et al. (2009), who found that $8 \pm 3$ per cent of the *SFR* is directly triggered by major merger in a similar range of mass and redshift (see also Jogee et al. 2009). This is a strict lower limit to the contribution of major mergers to the *SFR* density since it considers only a single phase of the merger process, while part of the pre-fusion and relaxation phase *SFR* can be also triggered by the merger event itself; the first passage between the two progenitors can also result in a small peak of *SFR* as suggested by hydrodynamical simulations (Cox et al. 2008; Cibinel et al. 2018; Silva et al. 2018), while part of the gas expelled during the merger can fall back on the remnant and reform a stellar disc (e.g. Robertson et al. 2006; Hopkins et al. 2009b).

In summary, there appears to be no contradiction between the important role that major mergers were suggested to play in the recent structural evolution of most local spiral galaxies and the cosmic evolution of the *SFR* density at $z < 1$.

## 3 DATA

### 3.1 Sample description

We used the IMAGES-CDFS sample of 33 star-forming galaxies at $z \sim 0.6$ with stellar masses in the range $M_{\rm stellar} = 10^{10-11}$ M$_\odot$.[4] This sample is a subsample of the IMAGES survey of $z = 0.4$–0.75 galaxies (Yang et al. 2008). The IMAGES-CDFS sample is fully described and characterized in P12, in which it was shown that it is representative of the intermediate-mass population at $z \sim 0.6$ (see Introduction). The spatially resolved kinematics (from FLAMES-GIRAFFE at VLT) and resulting kinematic classification of this sample was presented in Yang et al. (2008), while their HST high-resolution morphology was studied in Neichel et al. (2008). As further discussed in Section 5.3, this sample follows all classical scaling relations with a scatter similar to that found in larger samples: *SFR* versus $M_{\rm stellar}$ (the so-called main sequence of star-forming galaxies, see P14), rotation velocities $V_{\rm rot}$ versus $M_{\rm stellar}$ (the TFR, Puech et al. 2010), $j_{\rm disc}$ versus $M_{\rm stellar}$ (the Fall relation), size versus $M_{\rm stellar}$, or between the gas velocity dispersion $\sigma_{\rm gas}$ and *SFR* (see Fig. 7).

The morpho-kinematic observations of these galaxies were compared to a representative grid of major merger hydrodynamical models to identify merger candidates (Hammer et al. 2009b, hereafter H09). Morphological and kinematic maps from these simulations were produced and resampled to the spatial resolution of the observations. These maps were then explored visually as a function of time and viewing angle to determine the best match to observations. Secure major merger candidates were identified for the two-third of the sample (23 galaxies). The time at which the model provides the best match to morpho-kinematic observations was used to estimate in which phase of the merging process each galaxy was observed; the pre-fusion phase, in which the two progenitors can still be distinguished (requiring to identify the two nuclei if the progenitors are overlapping), the fusion phase, in which the two progenitors are actually merging, or the virialization phase, in which a disc can reform provided that (at first order) the gas fraction is high enough (see Section 2). Since star-forming galaxies account for 60 per cent of intermediate-mass galaxies at this redshift (Hammer et al. 1997), this led to 33 per cent of $z \sim 0.6$ intermediate-mass galaxies associated with any phase of major mergers with baryonic mass ratio $\geq 1{:}4$, and possibly up to $\sim$50 per cent including unsecured candidates. In what follows (except in Section 5.3), we restricted the sample to the 23 objects for which major merger models were securely identified by H09. The time evolution of the properties detailed in Section 3.2 along these 23 merging galaxies are shown in Figs 1–4, which represent the core of this study. The subsections below define the different quantities, how they were estimated and, if necessary, normalized.

### 3.2 Measured and estimated physical quantities

#### 3.2.1 Star formation rate SFR

The total ($UV + IR$) *SFR*s were estimated in Puech (2010), to which we refer for details, with a resulting average uncertainty $\sim$0.12 dex. Because *SFR* depends on the total mass of gas [as reflected by the Schmidt–Kennicutt (SK) relation, e.g. Kennicutt & Evans 2012], the measured *SFR*s were homogenized by simply dividing the observed *SFR* by the estimated total mass of gas in each galaxy (see below) and multiplying each derived number by the average mass of gas in the sample. This normalized *SFR* is referred to as *SFR*$^*$ in the following. This avoids possible biases when comparing galaxy *SFR*s with significantly different gas reservoirs.

#### 3.2.2 Gravitational potential perturbation $\Delta\phi/\phi$

Most galaxy mergers in the considered range of mass are expected to virialize towards spiral galaxies (see Section 2). The dynamics of the remnants is therefore expected to resemble that of a rotating disc just after several Gyr (e.g. Hopkins et al. 2009b). In Puech et al. (2008), the velocity fields (VF) of all galaxies in the sample were fitted by rotating disc models, and the residuals were used to build an estimate of the relative state of perturbation of the galaxy gravitational potential $\Delta\phi/\phi$, as described in P12.

#### 3.2.3 Half-light radius $R_{\rm half}$

The half-light radii of the galaxies were estimated in Neichel et al. (2008) with typical uncertainties $\sim$0.1 kpc. The observed $z$ band was chosen so that the derived half-light radius $R_{\rm half}(z)$ is representative of the stellar phase extent since it samples rest-frame light redwards the 4000 Å break. To avoid possible biases due to the expected increase of size as a function of stellar mass, we corrected $R_{\rm half}(z)$ using the relation measured by Ichikawa, Kajisawa & Akhlaghi (2012) for star-forming galaxies at $0.5 \leq z \leq 0.75$, which scales as $R_{\rm half} \propto M_{\rm stellar}^{0.119}$.

#### 3.2.4 Velocity dispersion $\sigma_{\rm gas}$

The average gas velocity dispersion in the outer regions of the galaxies were improved compared to the original method described in Puech et al. (2007). Beam smearing effects were corrected following the method proposed by Stott et al. (2016), i.e. by quadratically correcting the measured velocity dispersions (quadratically corrected for instrumental broadening) in each spaxel by the maximal velocity gradient between a given spaxel and all surrounding spaxels. The spatial average of the velocity dispersion was estimated as the S/N-weighted mean of the spaxels, excluding the central one to limit contamination by beam smearing in the central region. The resulting value was corrected of the mean expected effect of inclination on the combination of the three velocity components as described in Puech et al. (2007). The resulting median uncertainty on $\sigma_{gas}$ is $\sim 2$ km s$^{-1}$.

---

[4]Stellar masses were derived in Puech et al. (2008) by fitting specific Bruzual & Charlot (2003) stellar population models to colours constructed using apparent magnitudes that are the closest to the rest-frame $J$ band to minimize k-correction uncertainties.

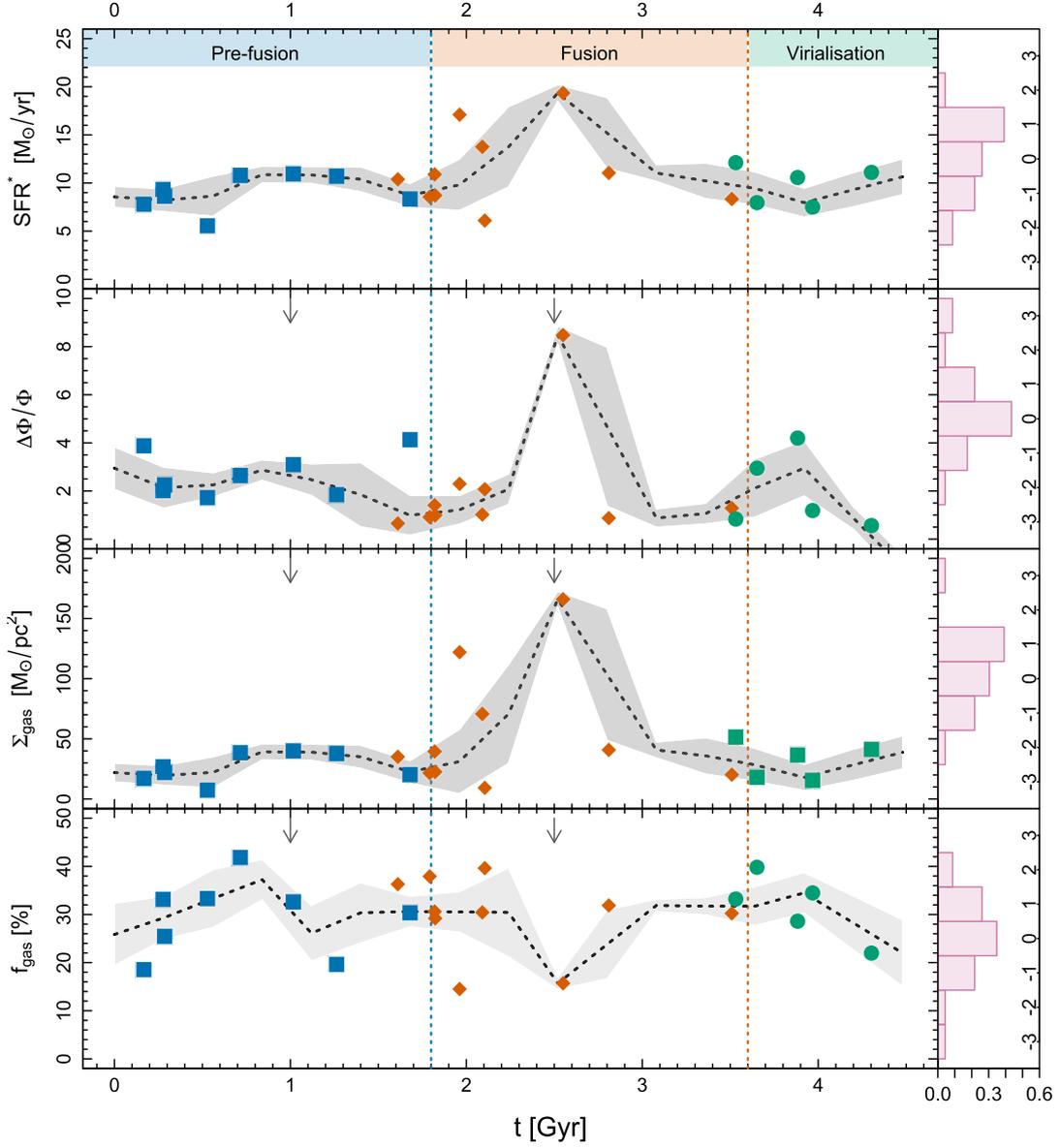

**Figure 1.** Evolution of the main properties related to star formation along the 23 galaxies representing an average major merger at $z \sim 0.6$. *Left-hand panels, from top to bottom:* (a) Time evolution of the normalized star formation rate ($SFR^*$), (b) variations in the gravitational potential $\Delta\Phi/\Phi$, (c) gas surface density $\Sigma_{gas}$, and (d) gas fraction $f_{gas}$ along an average major merger at $z \sim 0.6$. The median trend is shown as a dashed line, with associated uncertainty is grey. The blue dots represent galaxies in the pre-fusion phase, red dots galaxies in the fusion phase, and green dots galaxies in the virialization phase (see Section 4.1). The two downward arrows on the top of each panel indicate the position of the two $SFR^*$ peaks (see Section 4.2). When a physical quantity was normalized to correct for redshift evolution, this correction is indicated on the top of the panel (see Section 3.2). *Right-hand panels:* histograms of the distance between the observed points and the median trend, normalized by the associated uncertainty (see the text).

Several studies have noticed that velocity dispersion measurements in high-$z$ galaxies could still be significantly affected by residual beam smearing effects, resulting in overestimating the true velocity dispersion (e.g. Epinat et al. 2010; Davies et al. 2011). However, here the ratio between the galaxy disc scale length and the half-width half-maximum of the typical PSF during the observations $\sim 1$, which means that any possible systematic effect is probably $\leq 5$ per cent (see appendix B of Johnson et al. 2018), and can therefore be neglected.

### 3.2.5 Specific disc angular momentum $j_{gas}$

A forward modelling of the observed datacubes was used to estimate the beam-smearing and inclination-corrected rotation velocities $V_{rot}$ in Puech et al. (2008). These can be used to estimate the (gas) specific angular momentum $j_{gas} = 2R_d V_{rot}$, in which $R_d$ is the disc scalelength, which can be estimated using $R_d = R_{half}/1.68$ for a simple exponential disc. More elaborated estimators were defined over the past years but the simplified exponential approximation appears to be accurate within 0.1–0.15 dex (Romanowsky & Fall 2012; Obreschkow & Glazebrook 2014), which remains within the individual uncertainties on $V_{rot}$ and $R_{half}$ (both propagating in a median uncertainty on $j_{gas} \sim 0.1$ dex) and is therefore sufficient for the present purpose. To normalize by the average relations measured between size versus mass and velocity versus mass in galaxies, we corrected the estimated $j_{gas}$ using the scaling relations $R_{half} \propto M_{stellar}^{0.119}$ (see above) and $V_{rot} \propto M_{stellar}^{0.357}$, the latter being inferred from the TFR measured in the same sample (Puech et al. 2008, 2010).

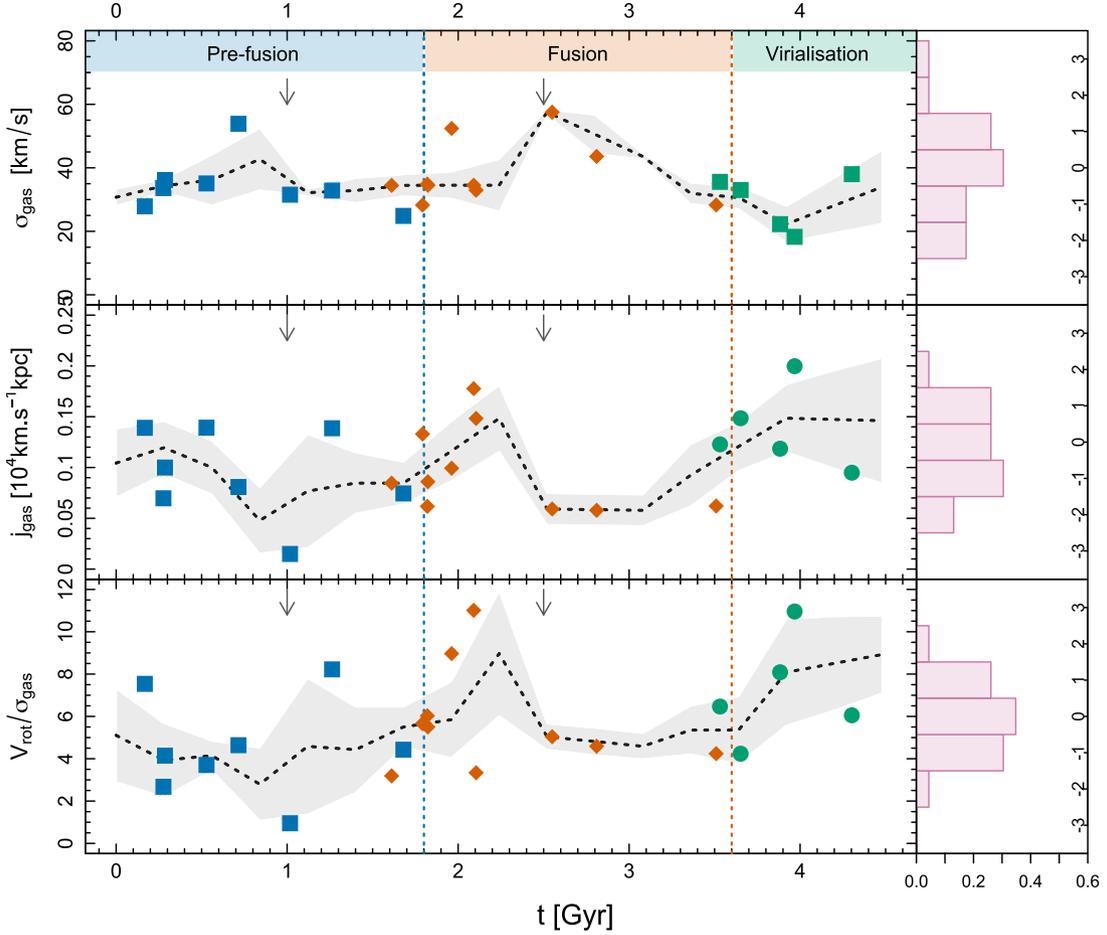

**Figure 2.** Evolution of the main dynamical properties along the 23 galaxies representing an average major merger at $z \sim 0.6$. *Left-hand panels, from top to bottom:* (a) Time evolution of the velocity dispersion $\sigma_{gas}$, (b) (normalized) gas specific angular momentum $j_{gas}$, and (c) gas dynamical support $V_{rot}/\sigma_{gas}$ along an average major merger at $z \sim 0.6$. Symbols are the same as that in Fig. 1; in particular, the two downward arrows on the top of each panel indicate the position of the two *SFR*$^*$ peaks (see Section 4.2). *Right-hand panels:* histograms of the distance between the observed points and the median trend, normalized by the associated uncertainty (see the text).

### 3.2.6 Dynamical support $V_{rot}/\sigma_{gas}$

The dynamical support of each galaxy was estimated following Puech et al. (2007) as the simple ratio between the estimated gas rotation velocity and average gas velocity dispersion (see above). Both $V_{rot}$ and $\sigma_{gas}$ were corrected for inclination (on a one-by-one basis for $V_{rot}$ and statistically for $\sigma_{gas}$, see above) so this quantity should be largely independent of projection effects and of mass variations in the sample.

### 3.2.7 Gas surface density $\Sigma_{gas}$

The SK relation between the total (atomic + molecular) gas surface density to the *SFR* surface density (e.g. Kennicutt & Evans 2012) can be used to estimate the gas surface density $\Sigma_{gas}$ from the observed *SFR* surface density $\Sigma_{SFR}$ in distant galaxies in which no direct H I or CO measurements are available, which was done in Puech et al. (2010) for the present sample. However, it has been suggested that the above canonical SK relation might be steeper in strong starbursts such as (U)LIRGs (Bouché et al. 2007; Daddi et al. 2010). In the range of $\Sigma_{SFR}$ considered here ($\sim 0.004$–$0.46\,M_\odot$ yr$^{-1}$ kpc$^2$), this might translate in systematically smaller $\Sigma_{SFR}$ in LIRGs by $\sim 0.33$ dex (e.g. Bouché et al. 2007). We did not attempt to use separate SK relations for normal and starbursts/LIRGs galaxies given that this systematic effect is comparable to the measured scatter of the SK relation (Kennicutt & Evans 2012).

### 3.2.8 Gas fraction $f_{gas}$

Gas masses $M_{gas}$ were estimated from the total gas surface densities $\Sigma_{gas}$ and total sizes (see Puech et al. 2010). Gas fractions were then simply derived as $f_{gas} = M_{gas}/(M_{gas} + M_{stellar})$.

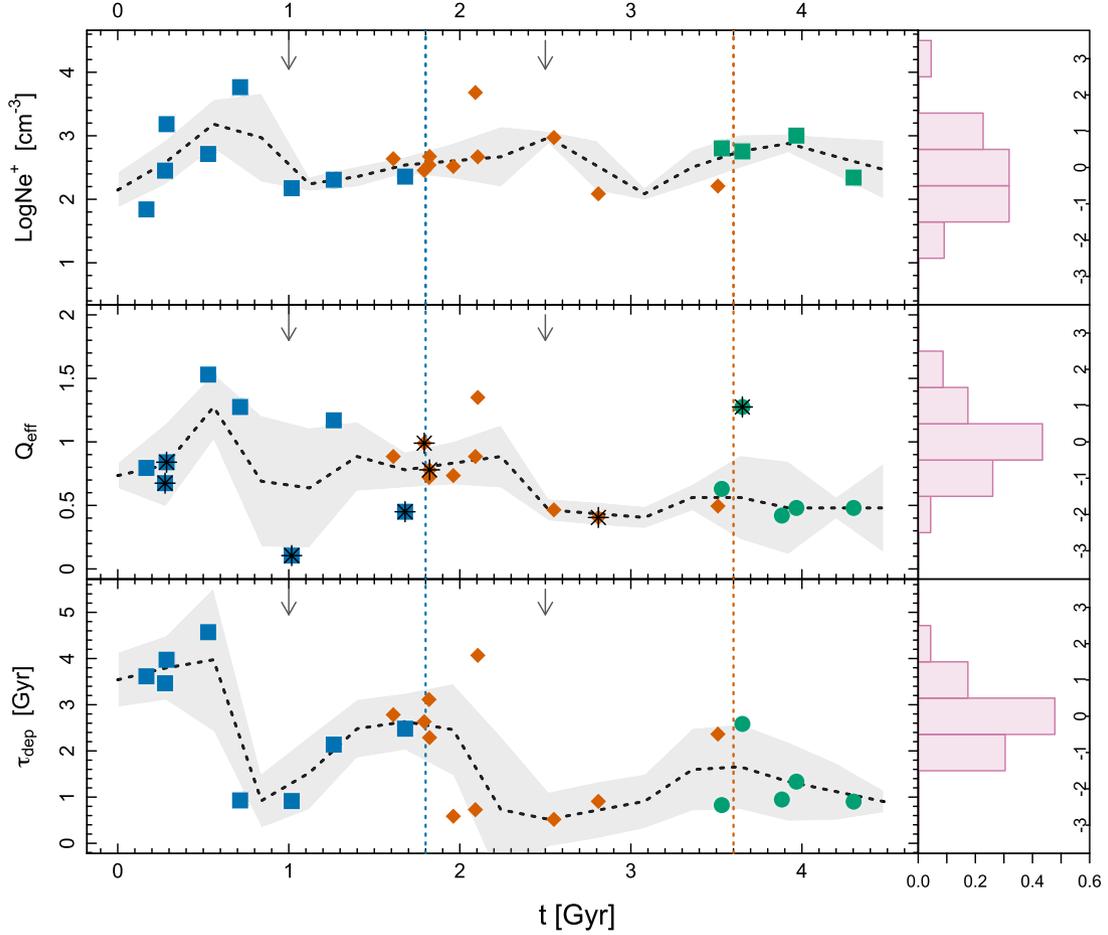

**Figure 3.** Evolution of ISM properties along the 23 galaxies representing an average major merger at $z \sim 0.6$. *From top to bottom:* (a) Time evolution of the electron density in the densest regions $N_e^+$, (b) effective Toomre stability factor $Q_{\rm eff}$ with clumpy galaxies indicated using superimposed star symbols, and (c) total gas depletion time-scale $\tau_{\rm dep}$ along an average major merger at $z \sim 0.6$. Symbols are the same as that in Fig. 1; in particular, the two downward arrows on the top of each panel indicate the position of the two *SFR*$^*$ peaks (see Section 4.2). *Right-hand panels:* histograms of the distance between the observed points and the median trend, normalized by the associated uncertainty (see the text).

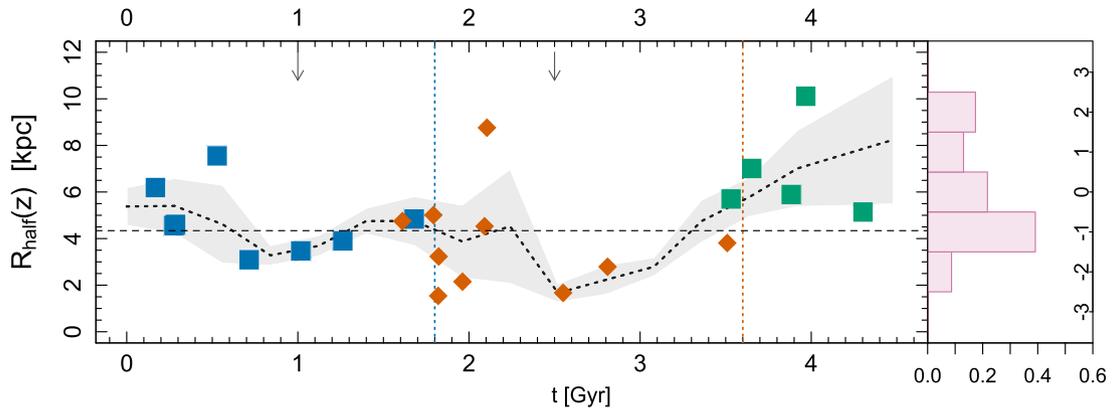

**Figure 4.** Size evolution along the 23 galaxies representing an average major merger at $z \sim 0.6$. *From top to bottom:* Time evolution of the (normalized) half-light radius $R_{\rm half}$ along the average $z \sim 0.6$ major merger. Symbols are the same as that in Fig. 1; in particular, the two downward arrows on the top of each panel indicate the position of the two *SFR*$^*$ peaks (see Section 4.2). *Right-hand panels:* histogram of the distance between the observed points and the median trend, normalized by the associated uncertainty (see the text).

*3.2.9 Electron density* $N_e$

Electron density maps were constructed from the FLAMES-GIRAFFE datacubes as detailed in Puech et al. (2006b): electron densities $N_e$ were estimated from the [O II] $\lambda3726/\lambda3729$ line ratio using the Temden task of IRAF (Shaw & Dufour 1995), and assuming an electron temperature of 10 000 K, which is typical of non metal-poor or metal-rich H II regions (Osterbrock & Ferland 2006). A low signal-to-noise ratio on the [O II] emission line can easily translate into a large uncertainty on the derived electron density, especially at very low/high $N_e$ since the calibration between the line ratio and $N_e$ then saturates (see Osterbrock & Ferland 2006). To limit these uncertainties, the electron density maps were thresholded using the [O II] emission line flux signal-to-noise ratio $S/N$ to keep only spaxels with $S/N \geq 5$. Then we considered for each galaxy the mean of the 50 per cent highest $N_e$ values to trace the typical electron density in the densest regions, which we refer to as $N_e^+$.

Sanders et al. (2016) noticed that the collision strength and transition probabilities used in Temden are outdated and they conducted a comparison with updated values (see their fig. 1). The new and the old calibrations give similar values at $\sim 100$ cm$^{-3}$ (which corresponds to the mean value of most galaxies in the sample), while the new calibration gives larger values above, typically by a factor 2 at $\sim 1000$ cm$^{-3}$ and larger above. Since we will consider trends in the densest regions as traced by $N_e^+$, using the IRAF calibration will actually provide underestimated values. As we will see below, $N_e^+$ is found to be relatively constant around $\sim 250$ cm$^{-3}$ in the sample so that the effect of a new calibration should not affect the median trend discussed below apart for a moderate systematic shift upwards.

*3.2.10 Toomre effective stability factor* $Q_{eff}$

The Toomre effective stability factor $Q_{eff}$ was derived as in Puech (2010) following the approximation proposed by Wang & Silk (1994), i.e. combining contributions from gas and stellar phases. Romeo & Wiegert (2011) studied in detail the accuracy of this approximation and they found that $Q_{eff}$ can be largely underestimated if the stabilizing effect of disc thickness is not taken into account. They introduced a correcting factor $T \sim 0.8 + 0.7 \times (\sigma_z/\sigma_R)$, in which $\sigma_z$ and $\sigma_R$ are the vertical and radial velocity dispersions, respectively. In the present sample, the $\sigma_z/\sigma_R$ ratio is likely $\sim 1$ for both gas and stars, since gas is collisional and stars have not formed long enough to make the velocity ellipsoid significantly anisotropic. We therefore assumed that $T_{gas} = T_{stars} = 1.5$. The effect of disc thickness can then be accounted for by simply substituting $T.Q_{eff}$ to $Q_{eff}$.

*3.2.11 Gas depletion time-scale* $\tau_{dep}$

The gas depletion time-scale was estimated as $\tau_{dep}[\text{Gyr}] = M_{gas}[\text{M}_\odot]/SFR[\text{M}_\odot\,\text{yr}^{-1}] \times 10^{-9}$, in which $SFR$ is the total star formation rate and $M_{gas}$ is the total gas mass. We emphasize that $\tau_{dep}$ is here defined as the *total* gas depletion time-scale, which is larger than the *molecular* depletion time-scale widely studied at high redshifts (e.g. Daddi et al. 2010; Scoville et al. 2016).

### 3.3 Constructing time-averaged trends

Once all the above quantities were estimated for each galaxy, these were sorted as a function of time along the merging process. For this, we used the time-scales derived in H09, which correspond to the times at which the best matches between the observations and the grid of major merger models were found (see coloured dots in Figs 1–4). When relevant, the normalization applied to the estimated quantities is indicated in the different panels.

Time-averaged trends were constructed using a running median with a time resolution $t_{resol} = 0.28$ Gyr, which corresponds to twice the median time step between successive pairs of observed points. At each point along this grid (which ranges from 0 to 4.48 Gyr), we median-averaged all the observed points within $\pm t_{resol}$. The resulting median trends are shown as black dashed lines in Figs 1–4. Uncertainties on each median trend were derived using bootstrap resampling and convolving the r.m.s. value obtained considering the complete sample with the r.m.s. value obtained in the samples drawn from each $\pm t_{resol}$ interval, to account for both scatters at long and short time-scales. The last point of the median grid slightly extended beyond the maximal time sampled by observations ($t = 4.3$ Gyr), and was extrapolated linearly from the previous point.

This method assumes that ensemble-average values can be estimated from time-average values. This is the reason why some of the estimated quantities were normalized by their measured evolution as a function of $z$, when such an evolution could impact the resulting time-average trend. Another complication can arise from the fact that some of the considered properties are particularly difficult to measure during the fusion phase. This is for instance the case for the specific angular momentum and more generally of all quantities that directly depend on rotation velocity, which is difficult to measure accurately during such a phase in which the two progenitor discs are destroyed and can be hardly disentangled, especially at the coarse spatial resolution of the FLAMES-GIRAFFE IFU observations ($\sim 7$ kpc). Therefore, such properties have to be interpreted carefully during this phase.

The resulting median trends are shown in the left-hand panels of Figs 1–4. In the right-hand panels we show the histogram of the distance between the observed points and the median trend, normalized by the uncertainty as determined above. These panels reveal that the observed quantities are all well accounted for by the associated uncertainties since all observed points are distributed within $\pm 3\sigma$ around the median trends.

## 4 TIME-AVERAGE PROPERTIES OF MAJOR MERGERS BETWEEN STAR-FORMING GALAXIES AT $z \sim 0.6$

### 4.1 Mergers initial conditions

The initial conditions for the best-fitting major merger models were studied in detail in H09 and P12. We summarize here the main conclusions for the sake of completeness.

The progenitor median stellar mass ratio in the was found to be $\mu_{stellar} = 3.03 \pm 0.34$ (see fig. 3 of H09). Theoretical expectations in major merger processes (Hopkins et al. 2010a) predict that such interactions induce a variation in the gravitational potential of the main progenitor that scales as $\Delta\phi/\phi \sim \mu_{baryon}$, in which $\mu_{baryon}$ is the baryonic mass ratio between the progenitors. In the range of stellar mass considered here, $\mu_{baryon} \sim \mu_{stellar}^{0.888}$ (Hopkins et al. 2010c) so that $\mu_{stellar} = 3.03$ translates into $\mu_{baryon} = 2.7$. P12 consistently found that in the most kinematically disturbed galaxies $\Delta\phi/\phi \sim 3.1 \pm 1.0$; the average spatial scale and amplitude of these perturbations are consistent with expectations from major mergers, but not with those from minor mergers, external gas accretion, or secular evolution (either clump or bar instabilities, see P12).

H09 found that the orbital parameters range from polar to direct orbits with prograde and retrograde spin orientations (see their fig.

5) but with a clear preference for inclined orbits with prograde–prograde spin orientations. They also extrapolated the gas fractions in the progenitors (60–80 per cent) and found values in agreement with estimates in the expected range of mass and redshift (Rodrigues et al. 2012). The galaxies in the sample were further classified into three merger phases: (1) the pre-fusion phase during which the two progenitors can still be identified as distinct components, (2) the fusion phase during which they coalesce, and (3) the relaxation phase in which the remnant progressively virializes. In all figures, we used blue, red, and green dots, respectively, to reflect this classification. P12 derived an average duration ∼1.8 Gyr for the pre-fusion and fusion phases, which matches well with binary merger simulations in the same range of mass and gas (see detailed comparison in P12). The average phase durations are indicated as blue and red coloured boxes in Fig. 1(a).

The H09 classification into different merger phases was used to estimate the major merger rate in massive galaxies at $z = 0.7$–1.5 (P12). The merger rate was found to decrease from $\sim 0.1\,\mathrm{Gyr}^{-1}$ at $z = 1.6$ to $\sim 0.05\,\mathrm{Gyr}^{-1}$ at $z = 0.7$ in agreement with other measurements in the same range of mass and redshift (P12), including more recent results from cosmological simulations (see e.g. fig. 9 of Rodriguez-Gomez et al. 2015 in the Illustris simulation) or from observations when using similar selection criteria (see e.g. Man et al. 2016).

### 4.2 Star formation along the average merging process

P12 and P14 studied the evolution of $SFR^*$ along an average $z \sim 0.6$ major merger. In Fig. 1(a) we replot $SFR^*$ as a function of merging time,[5] now restricted to the 23 objects with secured major merger models and with the median time-average trend overplotted (instead of a simple visual guideline as in P12 and P14). Within uncertainties, the median trend closely follows the observations, with two peaks at 1 and 2.5 Gyr after the beginning of the interaction,[6] when the two progenitors are still constituting a well-separated pair ($T = 0$). Based on morpho-kinematic data, H09 and P12 interpreted the 1 Gyr peak as the first passage, during which a modest enhancement of $SFR$ is produced, while the 2.5 Gyr peak corresponds to the fusion time during which the progenitors coalesce into a single object and gas is funnelled into the central regions (e.g. Di Matteo et al. 2007; Cox et al. 2008). The two peak enhancements and durations are roughly consistent with state-of-the-art simulations of binary major mergers that include explicit feedback (see e.g. Hopkins et al. 2013), cosmological zoom-in simulations (e.g. Sparre & Springel 2016), as well as from independent constraints inferred from the scatter of the $SFR$–$M_{stellar}$ relation (see Section 2; P14). During the virialization phase, the remnant is expected to progressively rebuild a disc (see Section 2.2) in which equilibrium is reached faster in the central regions compared to the outer parts.[7]

Hydrodynamical simulations of binary major mergers suggest that all interactions do not always result in a strong enhancement of star formation during fusion depending on the exact orbits, mass ratio, and gas fractions in the progenitors (Di Matteo et al. 2007). Simulations have shown that retrograde orbits (i.e. when the spin orientations of the progenitors are not aligned with the orbital angular momentum orientation) tend to result in weaker SFR enhancements because such orbits are less in resonant interactions with the orbital motion, which reduces the resulting internal tidal torques in the primary (Cox et al. 2008; Hopkins et al. 2009b). This in turn limits the loss of angular momentum of the gas in the disc, and, as a result, reduces star formation enhancements compared to prograde orbits. H09 studied the orbits of the merger progenitors for the sample shown in Fig. 1 and found a large majority (∼80 per cent) of prograde orbits. This is consistent with a significant $SFR$ average peak observed during both first passage (pre-fusion phase) and coalescence (fusion phase). The relatively extended $SFR$ tail that follows the fusion $SFR$ peak during ∼1 Gyr is also consistent with prograde orbits dominating the average over all orbits, and it could further suggest that significant feedback is at work in the merger remnant (Hopkins et al. 2013, see also Section 5.3).

Fig. 1(b) shows the time evolution of variations in the gravitational potential $\Delta\phi/\phi$ along the average merger. The median trend follows closely the $SFR^*$, in particular regarding the positions of the peaks. Note that these two quantities correspond to fully uncorrelated measurements since they were estimated from $IR + UV$ photometry and from the spatially resolved Doppler shifts of the [O II] emission lines compared to rotating disc models, respectively. The phased evolution of $SFR^*$ and $\Delta\phi/\phi$ therefore strongly supports that the $SFR$ enhancements are driven by the torque resulting from the passage of the secondary object (see Hopkins et al. 2009b fig. 6; Sauvaget et al. 2018). The modest increase of $\Delta\phi/\phi$ during the virialization phase might indicate that the remnant virialization is not that smooth a process from a dynamical point of view, or that other perturbing processes are occurring on top of the average major merger (e.g. more minor mergers, disc instabilities), which could vary from case to case.

In Fig. 1(c) is shown the evolution of the gas surface density $\Sigma_{gas}$. The median trend again reveals a phased evolution with $SFR^*$, with two coinciding peaks. These peaks emerge above $\Sigma_{gas} \sim 10$–$20\,\mathrm{M}_\odot\,\mathrm{pc}^{-2}$ consistently with the classic interpretation of the observed threshold at these values in the SK relation (Kennicutt & Evans 2012) as a limit above which star formation is effectively triggered in galaxies. $\Sigma_{gas}$ remains on average above this threshold during the virialization phase, which indicates that significant star formation still occurs during this phase, at a rate $\sim 10\,\mathrm{M}_\odot\,\mathrm{yr}^{-1}$ (see panel a).

Finally, Fig. 1(d) shows the evolution of the estimated gas fraction. The scatter on $f_{gas}$ is significantly larger reflecting the fact that it is indirectly estimated from the inversion of the SK relation (see Section 3.2). Nevertheless, the median trend suggests two decreases in $f_{gas}$ consistently phased with the peaks in $SFR^*$ and $\Delta\phi/\phi$, according to the above interpretation (see also Georgakakis, Forbes & Norris 2000; Cox et al. 2008; Athanassoula et al. 2016). The average gas fraction in the sample is ∼30 per cent, consistent with estimates in this range of mass and redshift (Rodrigues et al. 2012). The gas fractions during the virialization phase show values similar to those at the beginning of the merger event, which suggests that the gas reservoir is refilled rapidly (i.e. <0.5 Gyr) after fusion. There are several possible origins for this refurbished gas, which are (1) gas expelled during the merger event and then re-accreted as it falls back on to the remnant (as seen in simulations, e.g. Barnes 2002; Hopkins et al. 2009b; Aumer et al. 2014), (2) gas accreted from the surrounding hot halo (e.g. Moster et al. 2011; Athanassoula et al. 2016), (3) the continuous recycling of baryons from stellar mass-loss (e.g. Martig & Bournaud 2010), (4) galactic fountains resulting from feedback (e.g. Brook et al. 2011, 2012a; Übler et al. 2014), or gas infalling from intergalactic filaments (Governato et al. 2009; Martin et al. 2018). The non-evolution (or possibly weak evolution within uncertainties) of the baryonic TFR since $z \sim 0.6$–1.5 (Puech et al. 2010; Vergani et al. 2012) suggests that at least part of this gas was already gravitationally bound to the stellar discs of their progenitors.

---
[5]Since $SFR$s were derived using the rest-frame $UV$ and $IR$ photometry, the $SFR$ measurements are sensitive on time-scales ≤200 Myr (Kennicutt & Evans 2012), which implies that the $SFR$ temporal variations shown in Fig. 1(a) are limited by the adopted time step (0.28 Gyr) rather than the $SFR$ measurement.

[6]In all panels of Figs 1–4, we indicated the position of the two $SFR$ peaks by two downward arrows to ease the comparison between the different quantities and help assess how phased the evolutions between the different properties are.

[7]which can make the spatially resolved kinematics of the reforming disc appear relatively similar to that of a fully virialized disc (Robertson & Bullock 2008; Bellocchi, Arribas & Colina 2012).

### 4.3 Dynamical temporal evolution

Fig. 2(a) shows the evolution of the gas velocity dispersion $\sigma_{gas}$. An increase of $\sigma_{gas}$ is observed for each peak of $SFR^*$. These are consistent with the interaction converting gravitational energy into kinetic energy associated with gas turbulence that results from shocks produced during first passage and fusion. This interpretation is consistent with numerical models (Covington et al. 2010; Bournaud et al. 2011; Powell et al. 2013; Hung et al. 2015) and observations of local galaxies (Monreal-Ibero et al. 2010; Arribas et al. 2014; Rich, Kewley & Dopita 2015). It provides a physical explanation of the larger scatter of the TFR observed at high $z$ (Puech et al. 2010). It also explains why when using the tracer $S_{0.5} = (0.5 \times V_{rot}^2 + \sigma^2)^{-1/2}$ instead of $V_{rot}$, the scatter of the resulting TFR shrinks significantly at high $z$ (e.g. Kassin et al. 2007; Puech et al. 2010; Vergani et al. 2012). Since $S_{0.5}$ traces the total specific energy, it is therefore expected to be conserved at first order during the merging process (Covington et al. 2010).

The median temporal evolution of $j_{gas}$ reveals two decreases that are phased with the $SFR^*$ peaks (see Fig. 2b). During the pre-fusion and fusion phases, the gas probes mostly the inner regions that are expected to loose angular momentum as it is driven to the centre by a bar-like structure, which results from the torque exerted by the secondary progenitor as it passes by (see Hopkins et al. 2009b fig. 6). During the virialization phase, the gas is more sensitive to the re-accreted material that gained angular momentum from the orbital motion (see e.g. Übler et al. 2014 fig. 11), which results in a net, although moderate, gain $\sim 0.1$ dex, consistent with results of cosmological simulations for wet major mergers between star-forming galaxies in similar ranges of mass and redshift (Lagos et al. 2018; see also Sokołowska et al. 2017). The final $j_{gas}$ is consistent with measurements of galaxies of similar mass between $z = 0$ and $z = 1$ (Puech et al. 2007; Contini et al. 2016; Cortese et al. 2016; Harrison et al. 2017; Swinbank et al. 2017).

On the overall, the temporal evolution of $j_{gas}$ shows more scatter compared to the other properties, which is expected because of the simplified assumptions used here in regard of the expected morphological variations in the sample (see Section 3.2.5), and also because $j_{gas}$ can be a particularly difficult quantity to estimate from observations when the gravitational potential of the main progenitor is the most perturbed since the assumption of a thin exponential disc is then probably no more valid. The latter is probably reflected by the larger scatter in $j_{gas}$ during the pre-fusion and fusion phases, in which $\Delta\phi/\phi$ is also the largest (see Fig. 1b). In addition, as haloes evolve and interact they are expected to see their angular momentum varying as a non-linear random-walk process since it can increase or decrease depending on the exact mass ratio, gas fractions, orbits, geometry, dynamics, mass distributions, or feedback strength (see discussion in Puech et al. 2007), which should result in a large range of situations, hence in significant physical scatter during the virialization phase.

Fig. 2(c) shows that the strength of the rotational dynamical rotational support in the gaseous phase $V_{rot}/\sigma$ evolves in phase with $j_{gas}$. A decrease in $V_{rot}/\sigma$ is expected during the most kinematically perturbed phases of the merger (see also Puech et al. 2007). The first drop in $V_{rot}/\sigma$ reaches the limit of 3 that was used by Kassin et al. (2012) to define the fraction discs that are 'settled' as a function of mass and redshift. This threshold was also found to provide a good separation between 'rotating discs' and 'perturbed rotators' in the sample studied by Puech et al. (2007). However, measurements of $V_{rot}$ during the fusion phase are probably also affected by large errors (see above and Section 3.3), which prevent to relate $V_{rot}/\sigma$ to any absolute threshold related to 'disc settling' during this specific phase. The dynamical support in the gas eventually increases to $V_{rot}/\sigma_{gas} \sim 10$, which suggests that the merger remnants are virializing towards thin discs dynamically similar to local spiral galaxies (Epinat et al. 2010).

### 4.4 Temporal evolution of further ISM-related properties

Fig. 3(a) shows the temporal evolution of the electron density in the densest regions $N_e^+$ along the average merger process. We found $N_e^+ \geq 100\,\mathrm{cm^{-3}}$ with most values in the range 100–500 cm$^{-3}$, as found in H II regions of local interacting galaxies (Krabbe et al. 2014). The $N_e^+$ floor of $\sim 100\,\mathrm{cm^{-3}}$ found along the merging process is also consistent with the typical values reported in H II regions of local isolated galaxies (Krabbe et al. 2014). Hydrodynamical simulations suggest that during the peak of star formation activity, more ionized gas is channelled to high densities $\gtrsim 10^4\,\mathrm{cm^{-3}}$ (Hopkins et al. 2013, fig. 5). Fig. 3(a) suggest that, within the length of the adopted time step (0.28 Gyr), $N_e^+$ could indeed increase up to $\sim 1000\,\mathrm{cm^{-3}}$ and above during such phases, tracing shocks in the gas induced by the interaction. Although the median trend in $N_e^+$ remains significantly hampered by the large uncertainty associated with the measurement of $N_e$, this is consistent with observations of local mergers, in which shock ionization in the outer disc regions appears to increase in more advance merger phases (Monreal-Ibero et al. 2010; Rich et al. 2015).

Fig. 3(b) shows the temporal evolution of $Q_{eff}$. For a gravitational disc to fragment into clumps, it is usually required that $Q_{eff} < 1$, which means that the system is unstable to both radial and axisymmetric perturbations (e.g. Polyachenko, Polyachenko & Strel'Nikov 1997; Griv 2006; but see Inoue et al. 2016; Oklopčić et al. 2017). We also found two minima corresponding to the two $SFR^*$ and $\Delta\phi/\phi$ enhancement peaks. We identified in the sample the galaxies that were classified as clumpy galaxies by Puech (2010) (see Fig. 3b), and found that these are indeed distributed along the whole merging sequence with a preference for clumpy disc to have the smallest $Q_{eff}$ values. However, the majority of clumpy galaxies appear to be located out of the two $Q_{eff}$ minima. This is consistent with a significant fraction of clumpy galaxies at $z \sim 0.6$ triggered by gravitational interactions (Puech 2010), but not necessarily by the phases corresponding to the strongest gravitational perturbations, in which the increased $\sigma_{gas}$ values can partly counterbalance the strong $\Sigma_{gas}$ peaks and re-stabilize the discs for a few 100s of Myr (see Fig. 2a). From the fraction of clumpy galaxies among the sample in the pre-fusion and fusion phases, we can estimate the average visibility time-scale during which galaxies are within a clumpy phase $\sim 1.2 \pm 0.4$ Gyr, consistent with simulations (Oklopčić et al. 2017). However, the lifetime of individual clumps is likely to be one order of magnitude smaller, with still some debate about whether clumps are short-lived (with lifetimes <100 Myr, e.g. Genel et al. 2012; Hopkins et al. 2012; Oklopčić et al. 2017) or long-lived (with lifetimes >100 Myr, e.g. Ceverino, Dekel & Bournaud 2010; Ceverino et al. 2012; Bournaud et al. 2014; Mandelker et al. 2017).

The evolution of the total gas depletion time-scale $\tau_{dep}$ also reveals two minima (see Fig. 3c), which are again well-phased with the two $SFR^*$ peaks, as expected (e.g. Di Matteo et al. 2007; Sparre & Springel 2016). In the pre-fusion phase, values are typical of those found at roughly half the optical radius of local discs (Leroy et al. 2008), or in galaxies with companions (Knapen & James 2009). It decreases along the average major merger sequence, reaching two minima at ∼1 Gyr, which indicate a fast consumption of the gas into stars typical of starbursts (Knapen & James 2009; Schiminovich et al. 2010; Jaskot et al. 2015), before stabilizing at ∼1–2 Gyr in the virialization phase. These values are consistent with the depletion time-scales found in post-starburst galaxies extracted from cosmological simulations, which also found that major mergers are an important channel for forming these objects at redshifts similar to those considered in this study (Davis et al. 2019). The low $\tau_{dep}$ values found in the virialization phase indicate that the gas reservoir would be quickly exhausted without any gas refurbishment, as previously discussed in Section 4.2.

### 4.5 Morphological evolution

In Fig. 4(a) we show the evolution of the *z*-band half-light radius $R_{half}(z)$. During the virialization phase, $R_{half}(z)$ increases by ∼50 per cent, which reflects an inside-out formation of stars in the ongoing disc rebuilding (González Delgado et al. 2015). In this last phase, $R_{half}(z)$ shows values consistent with measurements in local late-type galaxies at similar rest-frame wavelengths and stellar mass (∼5–8 kpc see e.g. Lange et al. 2015; Roy et al. 2018).

As expected (Hammer et al. 2001; Lotz et al. 2010), a compact phase is seen both during first passage and fusion. The dotted line represents the limit for compactness adopted in Hammer et al. (2001) to study luminous compact galaxies (LCGs). This limit corresponds to the ∼50 per cent smallest galaxies at the median mass of the sample ($\log(M_{stellar}/M_\odot) \sim 10.3$, see van der Wel et al. 2014). LCGs represent ∼20 per cent of the galaxies between $z = 0.4$ and 1 (Zheng et al. 2004) and contribute 40–50 per cent of the increase in the cosmic star formation history between $z = 0$ and 1, as measured from the rest-frame UV luminosities (i.e. unobscured star formation; Guzmán et al. 1997). Their spatially resolved kinematics revealed unvirialized motions for at least 32 per cent and possibly up to 82 per cent of them that are probably related to mergers (Puech et al. 2006a). The study of their local analogues reveal similar perturbed motions (e.g. Östlin et al. 2001, 2015), with a significant fraction of them (45 per cent in their selection criteria) showing a companion (Pérez-Gallego et al. 2011). According to the median trend in $R_{half}(z)$, 40 per cent (resp. 60 per cent) of the compact phase occurs during first passage (resp. fusion), in rough agreement with the results of Pérez-Gallego et al. (2011).

According to Fig. 4(a), the LCG phase is found to last 1.8 Gyr in total, which is consistent with demographics on the global population of LCGs between $z = 0.4$ and 1 in roughly the same range of stellar mass: since 19 per cent of them are found to be LCGs (Zheng et al. 2004), and that 35 per cent of local galaxies are expected to have undergone a major merger since $z = 1$ with progenitors in this range of mass (P12; see their fig. 5), one can roughly estimate that between $z = 0.4$ and $z = 1$ (corresponding to an elapsed time of 3.3 Gyr) the progenitors of these mergers have indeed spent $0.19 \times 3.3/0.35 = 1.8$ Gyr in an LCG phase.

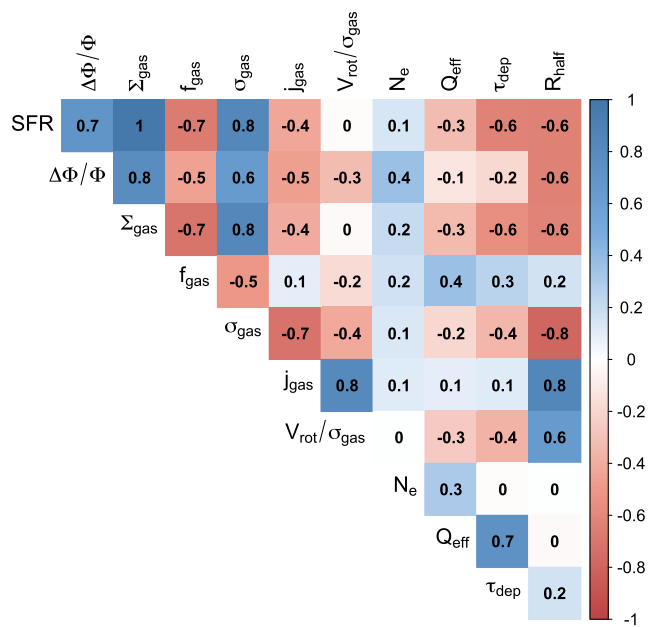

**Figure 5.** Correlation matrix between the different time-averaged quantities shown in Figs 1–4. The values correspond to the linear Pearson correlation coefficient.

## 5 DISCUSSION

### 5.1 Major mergers are on average driving SFR enhancements at $z \sim 0.6$

Figs 1–4 reveal that the evolution of the time-averaged quantities related to star formation, the ISM properties, dynamics, and morphology are globally consistent with expectations from major merger models or observations. We quantify trends in their co-evolution in Fig. 5, in which we show the cross-correlation coefficient matrix between all these quantities. This matrix supports quantitatively the following conclusions:

(i) There is a strong correlation (i.e. correlation coefficients ≥ 0.7) between the *SFR* evolution, the perturbations in the gravitational potential $\Delta\phi/\phi$, the gas surface density $\Sigma_{gas}$, and the gas velocity dispersion $\sigma_{gas}$.[8] This is interpreted as gas compression due to gravitational torques resulting from the major merger events, which in turn results in the increase of gas surface density that drives the *SFR* peaks during the pre-fusion and fusion phases. The gravitational energy injected during the interaction is converted into kinematic energy through shocks, which result into two timely phased enhancements of $\sigma_{gas}$;

(ii) The evolution of the gas velocity dispersion $\sigma_{gas}$ is strongly anticorrelated with the evolution of $j_{gas}$ and $R_{half}$; this reflects that on average specific angular momentum is lost during the most perturbed phases of the merger (also reflected in the strong correlation between $j_{gas}$ and dynamical support $V_{rot}/\sigma$; the system looses at the same time dynamical support from rotation). This translates into a compact phase consistent with the observed properties of LCGs;

(iii) Moderate correlations (i.e. correlation coefficients ∼0.4–0.7) with quantities more prone to uncertainties also support the

---

[8]Note that *SFR* and $\Sigma_{gas}$ are not independent quantities since they both depend on the same observable *SFR*. However, $\sigma_{gas}$, $\Delta\phi/\phi$, and *SFR* are fully independent.

Table 1. Median average trends as a function of time of the physical properties shown in Figs 1–4.

| Time [Gyr] | SFR* [M$_\odot$ yr$^{-1}$] | $\Delta\Phi/\Phi$ | $\Sigma_{\rm gas}$ [M$_\odot$ pc$^{-2}$] | $f_{\rm gas}$ [per cent] | $\sigma_{\rm gas}$ [km s$^{-1}$] | $j_{\rm gas}$ [10$^4$ km s$^{-1}$ kpc] | Log($N_e^+$) [cm$^{-3}$] | $Q_{\rm eff}$ | $\tau_{\rm dep}$ [Gyr] | $R_{\rm half}(z)$ [kpc] |
|---|---|---|---|---|---|---|---|---|---|---|
| 0.00 | 8.6 ± 1.1 | 2.9 ± 0.9 | 22.0 ± 8.0 | 0.26 ± 0.06 | 30.7 ± 2.6 | 0.10 ± 0.03 | 2.1 ± 0.3 | 0.7 ± 0.1 | 3.5 ± 0.6 | 5.4 ± 0.8 |
| 0.28 | 8.2 ± 1.3 | 2.1 ± 0.9 | 19.6 ± 8.5 | 0.29 ± 0.05 | 34.3 ± 2.0 | 0.12 ± 0.03 | 2.6 ± 0.4 | 0.8 ± 0.3 | 3.8 ± 0.7 | 5.4 ± 1.2 |
| 0.56 | 8.6 ± 2.1 | 2.3 ± 0.5 | 22.1 ± 13.0 | 0.33 ± 0.06 | 36.2 ± 8.1 | 0.10 ± 0.03 | 3.2 ± 0.4 | 1.3 ± 0.3 | 4.0 ± 1.5 | 4.6 ± 1.7 |
| 0.84 | 10.9 ± 0.9 | 2.9 ± 0.4 | 39.2 ± 6.9 | 0.37 ± 0.04 | 42.7 ± 9.7 | 0.05 ± 0.03 | 3.0 ± 0.7 | 0.7 ± 0.5 | 0.9 ± 0.6 | 3.3 ± 0.4 |
| 1.12 | 10.8 ± 0.9 | 2.5 ± 0.6 | 38.9 ± 6.9 | 0.26 ± 0.06 | 32.2 ± 1.1 | 0.08 ± 0.06 | 2.2 ± 0.1 | 0.6 ± 0.5 | 1.5 ± 0.8 | 3.7 ± 0.4 |
| 1.40 | 10.4 ± 1.3 | 1.8 ± 1.3 | 35.1 ± 9.8 | 0.30 ± 0.06 | 32.8 ± 4.0 | 0.08 ± 0.03 | 2.4 ± 0.2 | 0.9 ± 0.3 | 2.5 ± 0.6 | 4.8 ± 0.5 |
| 1.68 | 8.7 ± 1.3 | 1.0 ± 0.9 | 22.6 ± 9.4 | 0.31 ± 0.03 | 34.5 ± 3.5 | 0.08 ± 0.02 | 2.5 ± 0.1 | 0.8 ± 0.1 | 2.6 ± 0.6 | 4.8 ± 1.0 |
| 1.96 | 9.8 ± 2.7 | 1.2 ± 0.6 | 31.0 ± 27.3 | 0.31 ± 0.04 | 34.5 ± 4.5 | 0.12 ± 0.03 | 2.6 ± 0.3 | 0.8 ± 0.2 | 2.5 ± 0.9 | 3.9 ± 1.5 |
| 2.24 | 13.8 ± 4.3 | 2.1 ± 0.6 | 70.7 ± 40.9 | 0.30 ± 0.09 | 34.5 ± 8.2 | 0.15 ± 0.03 | 2.7 ± 0.5 | 0.9 ± 0.3 | 0.7 ± 1.6 | 4.5 ± 2.5 |
| 2.52 | 19.3 ± 0.9 | 8.5 ± 0.4 | 166.0 ± 6.9 | 0.16 ± 0.01 | 57.5 ± 0.9 | 0.06 ± 0.02 | 3.0 ± 0.1 | 0.5 ± 0.1 | 0.5 ± 0.6 | 1.7 ± 0.4 |
| 2.80 | 15.2 ± 3.7 | 4.7 ± 3.3 | 103.4 ± 53.8 | 0.24 ± 0.07 | 50.6 ± 6.1 | 0.06 ± 0.02 | 2.5 ± 0.4 | 0.4 ± 0.1 | 0.7 ± 0.6 | 2.2 ± 0.6 |
| 3.08 | 11.0 ± 0.9 | 0.9 ± 0.4 | 40.8 ± 6.9 | 0.32 ± 0.01 | 43.6 ± 0.9 | 0.06 ± 0.02 | 2.1 ± 0.1 | 0.4 ± 0.1 | 0.9 ± 0.6 | 2.8 ± 0.4 |
| 3.36 | 10.2 ± 1.9 | 1.1 ± 0.4 | 35.8 ± 15.0 | 0.32 ± 0.02 | 31.9 ± 3.3 | 0.09 ± 0.03 | 2.5 ± 0.3 | 0.6 ± 0.1 | 1.6 ± 0.9 | 4.8 ± 0.9 |
| 3.64 | 9.4 ± 1.7 | 2.1 ± 1.2 | 28.4 ± 13.9 | 0.32 ± 0.04 | 30.7 ± 3.6 | 0.12 ± 0.02 | 2.8 ± 0.3 | 0.6 ± 0.3 | 1.7 ± 0.9 | 5.8 ± 0.8 |
| 3.92 | 8.0 ± 1.5 | 2.9 ± 1.2 | 18.0 ± 10.9 | 0.34 ± 0.04 | 22.3 ± 5.7 | 0.15 ± 0.03 | 2.9 ± 0.1 | 0.5 ± 0.4 | 1.3 ± 0.8 | 7.0 ± 1.6 |
| 4.20 | 9.3 ± 1.8 | 0.9 ± 0.4 | 28.4 ± 13.0 | 0.28 ± 0.06 | 28.1 ± 8.6 | 0.15 ± 0.05 | 2.7 ± 0.3 | 0.5 ± 0.1 | 1.1 ± 0.6 | 7.6 ± 2.2 |
| 4.48 | 10.6 ± 1.9 | −1.2 ± 0.6 | 38.8 ± 13.6 | 0.22 ± 0.07 | 34.0 ± 11.4 | 0.15 ± 0.06 | 2.5 ± 0.5 | 0.5 ± 0.4 | 0.9 ± 0.3 | 8.2 ± 2.8 |

| | $V_{\rm rot}/\sigma_{\rm gas}$ |
|---|---|
| | 5.1 ± 2.2 |
| | 3.9 ± 1.8 |
| | 4.1 ± 0.7 |
| | 2.8 ± 1.7 |
| | 4.6 ± 3.2 |
| | 4.4 ± 2.0 |
| | 5.5 ± 1.0 |
| | 5.8 ± 1.8 |
| | 9.0 ± 3.0 |
| | 5.0 ± 0.6 |
| | 4.8 ± 0.7 |
| | 4.6 ± 0.6 |
| | 5.4 ± 1.1 |
| | 5.4 ± 1.6 |
| | 8.1 ± 2.5 |
| | 8.5 ± 2.2 |
| | 8.9 ± 1.8 |

above picture, e.g. the evolution of the gas fraction $f_{\rm gas}$ is moderately anticorrelated with the evolution of *SFR* (or $\Sigma_{\rm gas}$), $N_e^+$ is moderately correlated with $\Delta\phi/\phi$, while the Toomre instability factor $Q_{\rm eff}$ appears to be correlated with the gas depletion time-scale $\tau_{\rm gas}$, which further illustrates the intimate relation between dynamics and the rapid transformation of gas into stars.

The time step adopted (0.28 Gyr) does not allow to measure precisely any time delay between the evolution of all these quantities. We nevertheless note that in a more general study of star-forming galaxies (i.e. not necessarily merging), Hung et al. (2019) found in simulations that enhancements in velocity dispersion tend to coincide with increases of *SFR* with a typical time delay ∼50 Myr (see also Orr et al. 2018), which is consistent with the above results. While other processes such as gas accretion or minor mergers can also drive enhancements in *SFR*, Figs 1–4 and Fig. 5 evidence that major mergers result *on average* in a significant enhancements of *SFR* at $z < 1$. This does not imply that all *individual* major mergers (depending on orbits and gas fractions) are driving a significant peak of star formation (e.g. Di Matteo et al. 2007), or that the properties of the ISM in galaxies at $z > 1$ do not strongly reduce the enhancement in *SFR* as suggested by Fensch et al. (2017).

To ease the comparison between models and observations, we provide in Table 1 the median time-averaged evolution of all the properties shown in Figs 1–4.

### 5.2 Consequences for the morphology of present-day galaxies

The IMAGES survey, from which the present sample of merging galaxies was extracted, was initially designed to observe a representative sample of direct progenitors of local discs at $z \sim 0.6$ (Yang et al. 2008, see Introduction). Their observed morpho-kinematic properties suggested ongoing gas-rich major mergers at different phases (see H09; P12). Semi-empirical models showed that the global impact of such gas-rich mergers, integrated over look-back time, can statistically account for the demographics of bulges and discs along the local Hubble sequence (Hopkins et al. 2009a), including the reformation of large discs (Hopkins et al. 2009b). In this section, we investigate whether one can find a direct causal link between the *B/T* predicted from the morpho-kinematic properties of the IMAGES mergers combined with the physics of gas-rich major mergers, and the morphological distribution of galaxies along the local Hubble sequence.

The bulge-to-disc ratio *B/T* expected in merger remnants can be estimated (at zeroth order) following Hopkins et al. (2009b) as $B/T = \mu_{\rm b}^{-1} \times (1 - f_{\rm gas}^{\rm fusion})$,[9] in which $\mu_{\rm b}$ is the baryonic mass ratio between the progenitors and $f_{\rm gas}^{\rm fusion}$ is the gas fraction in the main progenitor at fusion time. The scatter in *B/T* that results from different possible orbital parameters is expected to be similar to that resulting from variations of $f_{\rm gas}^{\rm fusion}$ in the progenitors (Sauvaget et al. 2018). Here, we further assume that the orbital parameters of the mergers in the sample result in uniform scatter as a function of *B/T* (see also discussion in Hopkins et al. 2010c). The baryonic mass ratio $\mu_{\rm b}$ between the progenitors at fusion time were estimated from the stellar mass ratios $\mu_{\rm stellar}$ by interpolating the results of the abundance matching model tuned to the IMAGES sample (see Section 4.1). Finally, the gas fractions at fusion time $f_{\rm gas}^{\rm fusion}$ were estimated by correcting the estimated gas fractions in their progenitors (see H09) by the amount of gas consumed (or to be consumed) during the time that separates each ongoing merger observed at $z \sim 0.6$ to the average fusion peak time (using Fig. 1a and *SFR* measurements, see Section 3.2.1).

---

[9]Here, we adopted the convention $\mu = M_1/M_2$ where $M_1$ and $M_2$ and the most and less massive progenitors, respectively, while Hopkins et al. (2009b) used the opposite convention (i.e. $\mu = M_2/M_1$) in their original formula.

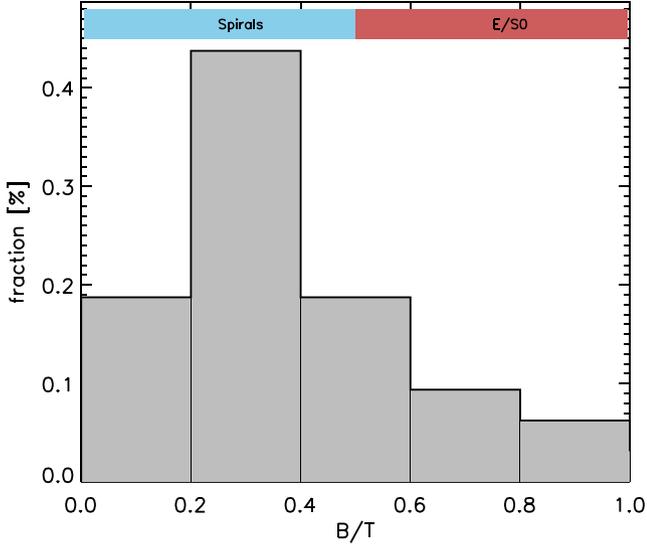

**Figure 6.** Distribution of *B/T* expected in the remnants of the gas-rich major mergers in the sample.

The resulting *B/T* distribution is shown in Fig. 6. The $z \sim 0.6$ gas-rich major mergers are expected to form preferentially $\sim$S0/a-Sb galaxies (see peak in Fig. 6 with average B/T $\sim$ 0.2–0.4[10]) but also a wide range of morphological late-type galaxies extending to Sbc (average B/T $\sim$ 0.1). Interestingly, García-Benito et al. (2017) found that the outer regions (i.e. between 1.5 and 2 highlight radius) of present-day spiral galaxies have stellar populations with ages $\sim$6 Gyr for galaxies in the range of expected stellar mass for the merger remnants (i.e. log($M_{stellar}$/ M$_\odot$) =10.3–11.1, see P12). A good match (within 1$\sigma$) between the distribution of the estimated ages in the outer regions of E to Sbc galaxies is also found with a particularly good match for S0 galaxies. We conclude that the distribution in expected morphological types for the merger remnants is consistent with the stellar population ages measured in local galaxies.

Adopting $B/T = 0.5$ as a simple limit between early-type (E/S0) and late-type (Sp) galaxies, we find a fraction of $66^9_{-7}$ per cent and $34^{10}_{-6}$ per cent of E/S0 and Sp, respectively, expected amongst the descendants of the present sample of star-forming galaxies (with uncertainties estimated using binomial statistics). Delgado-Serrano et al. (2010) found that at $z \sim 0.6$ star-forming and quiescent galaxies represent resp. 70 per cent and 30 per cent of the overall $z \sim 0.6$ galaxy population in the same range of mass (see their table 3). These results were used to compute the expected fractions of early (E/S0) and late-type (Sp) galaxies in Table 2. The total fraction of E/S0 and Sp at $z = 0$ is expected to be $\sim$68 per cent and 32 per cent, respectively, which matches very well the fractions determined at $z = 0$ from visual classifications (see Table 6 and Appendix B of P12 for details). Table 2 also illustrates that about half of local E/S0 and Sp should be remnants of relatively recent major mergers with the other half spared by any major interaction over the past $\sim$9 Gyr (z = 1.5), as consistently found by P12. This compares well with the prediction of Font et al. (2017), who found from cosmological simulations that 31 (54) per cent of local discs with masses similar to the MW were reformed after a major merger since $z = 1$ (2), respectively. Font et al. (2017) also found that 19 per cent of spheroids in the same range of mass have undergone a major mergers since $z = 1$, which also compares well with the fraction of 16 + 3 per cent deduced from Table 2. They also reported that 43 and 7 per cent of discs and spheroids at $z = 0$ are predicted to have escaped any major merger since z = 1. From Table 2, we find these fractions to be 31 and 13 per cent, respectively. Given the difference in the range of mass and redshift adopted by Font et al. (2017), and also the more restrictive limits adopted in their definition of discs,[11] we conclude that both observations and cosmological simulations provide a quantitatively coherent description of the impact of gas-rich major mergers on the morphology of present-day galaxies.

---

[10]We converted between *B/T* and morphological types using the results from Graham & Worley (2008).

[11]Font et al. (2017) defined as discs simulated galaxies with a kinematic disc-to-total ratio ≥ 0.3. This corresponds to a photometric $D/T \geq 0.7$ (Scannapieco et al. 2010), which in turn corresponds to the average *D/T* for Sa (Graham & Worley 2008), while we included within late-type galaxies S0/a galaxies, with an average $B/T = 0.2$ (or $D/T = 0.8$).

**Table 2.** Prediction of the $z \sim 0$ morphology (E/S0 versus Spirals determined according to their *B/T*) for the descendants of the IMAGES $z \sim 0.6$ galaxies. Fractions are given relative to the overall population at given *z*, while fractions indicated into brackets refer to the subpopulation of galaxies indicated in the directly above row.

| | | Overall $z \sim 0.6$ galaxy population | | | | | | Comments |
|---|---|---|---|---|---|---|---|---|
| | 70% of Star-Forming Galaxies | | | | 30% of Quiescent Galaxies | | | Hammer et al. (1997) |
| (31% of) | | (69% of) | (44% of) | | (31% of) | | (25% of) | Delgado-Serrano et al. (2010) |
| \| | | 48 ± 8% Pec | \| | | \| | 8 ± 2% Pec | \| | Merging galaxies (H09) |
| \| | (66% of) | (34% of) | \| | | \| | (66% of) | (34% of) | This study (see Fig. 6) |
| ↓ | ↓ | ↓ | ↓ | | ↓ | ↓ | ↓ | |
| 22 ± 6% Sp | 32 ± 8% Sp | 16 ± 4% E/S0 | 13 ± 2% E/S0 | | 9 ± 2% Sp | 5 ± 8% Sp | 3 ± 4% E/S0 | Descendants at z=0 |

| | Overall $z \sim 0$ galaxy population | | | Comments |
|---|---|---|---|---|
| | Sp | E/S0 | | |
| Prediction | 68% | 32% | | Total from above fractions |
| Nakamura et al. (2003) | 74 ± 3% | 28 ± 3% | | App. B of P12 |
| Nair & Abraham (2010) | 68% | 31% | | App. B of P12 |

### 5.3 Impact on the scatter of scaling relations

Figs 1–4 reveal significant temporal variations in several properties that might significantly enlarge the scatter of several scaling relations at high $z$ (see Section 2.4). Fig. 7 shows in the *full* IMAGES-CDFS sample (i.e. not restricted to mergers alone, see Section 3.1) some of the most discussed scaling relations in the literature, i.e. the so-called 'main sequence' of star-forming galaxies between *SFR* and $M_{stellar}$ (e.g. Noeske et al. 2007), the TFR between $V_{rot}$ and $M_{stellar}$ (e.g. Tiley et al. 2016), the relation between $\sigma_{gas}$ and *SFR*, the Fall relation between $j_{disc}$ and $M_{stellar}$ (Fall 1983; Contini et al. 2016; Harrison et al. 2017; Swinbank et al. 2017), and the relation between $R_{half}$ and $M_{stellar}$ (e.g. van der Wel et al. 2014).

Previous studies of the TFR at high $z$ revealed that the scatter of the relation increases significantly compared to $z \sim 0$ (e.g. Conselice et al. 2005). Spatially resolved kinematic surveys have shown that most (if not all) of this increase can be associated with galaxies with kinematic and/or morphological perturbations (Flores et al. 2006; Kassin et al. 2007; Puech et al. 2008, 2010). By isolating 'virialized rotating discs' (hereafter VRDs, see Rodrigues et al. 2017 for details) as galaxies that have both their morphological and kinematical spatial distributions consistent with expectations from local spiral galaxies,[12] the scatter is reduced from $\sim 0.8$ dex (in $\log(M_{stellar}/M_{\odot})$) to $\sim 0.15$ dex, i.e. similar to the local relation (Puech et al. 2010). Identifying galaxies that show both a morphology and a spatially resolved kinematics consistent with expectations from local spiral requires (1) both high-resolution spatial imaging and integral field spectroscopy and (2) a detailed analysis of the resulting data that are difficult to obtain and conduct within the samples of 100s of galaxies with spatially resolved kinematic measurements that are now assembled. This is one of the reason why the most recent kinematic surveys have adopted simpler criteria to select high-$z$ discs based on thresholds in $V_{rot}/\sigma_{gas}$ to use rotation-dominated galaxies as proxies (hereafter RDDs, see e.g. Kassin et al. 2012; Wisnioski et al. 2015; Tiley et al. 2016). While it is simpler and more directly related to the data to select high-$z$ discs as RDDs versus VRDs, the two criteria do not necessarily result into the same populations.

Fig. 7 compares the effect of selecting RDDs with $V_{rot}/\sigma_{gas} > 3$ (see e.g. Kassin et al. 2012; Tiley et al. 2016; see grey points in Fig. 7) versus VRDs (see black points in Fig. 7) on the scatter of the scaling relations. Both criteria appear to significantly reduce the scatter of all scaling relations involving kinematic measurements. However, both are relatively inefficient at reducing the scatter of other scaling relations such as the $R_{half}$ versus $M_{stellar}$ relation, which is expected since the latter does not directly imply any kinematic property. Note that both criteria are nevertheless able to remove the points scattered above the 'main sequence', which are well known to correspond to ongoing mergers (e.g. P14, Cibinel et al. 2018).

Fig. 7 reveals that VRDs are significantly more efficient in reducing the scatter of both the Tully-Fisher and Fall relations, which reduces from 0.3 to 0.05 dex in $\log(V_{rot})$ and from 0.3 to 0.2 dex in $\log(j_{disc})$, respectively. Note that both relations once restricted to VRDs show scatters similar to the local relations, which suggests that selecting high-$z$ discs as VRDs can result in scaling relations with most of measurement scatter removed. Selecting high-$z$ discs as RDDs, even adopting a relatively strict criterion with $V_{rot}/\sigma_{gas} > 3$, is not as efficient at removing this scatter. A closer look reveals that the relation defined by VRDs is shifted upwards in $\log(j_{disc})$ by $\sim 0.2$ dex compared to the relation defined by RDDs; this roughly corresponds to the evolution in the Fall relation between $z \sim 0.9$ and $z \sim 0$ as reported by Harrison et al. (2017, see also Swinbank et al. 2017). While it remains beyond the scope of this paper to study the evolution of the Fall relation, it suggests that selecting high-$z$ discs as RDDs can also hide systematic shifts within the residual scatter left in some of the scaling relations.

In the $\sigma_{gas}$ versus *SFR* plane, selecting high-$z$ discs as RDDs does not significantly decrease the scatter, and selecting them as VRDs perhaps slightly reduces the scatter, if anything. We also show in this plane typical predictions from the feedback-driven (see dark grey area) and the gravity-driven (see dashed lines) models proposed by Krumholz & Burkhart (2016). We followed their recommendations (see also Johnson et al. 2018) in the assumed fiducial values for the different model parameters, and show model prediction considering the median $V_{rot}$ in the sample. This means that most of the variations in the model predictions are expected to be due to variations in $Q_{gas}$ and $f_{gas}$, respectively; the dotted line corresponds to a fiducial value $Q_g = 1$, while the associated scatter encompasses a range $Q_g = 0.5$–2 (see grey dotted lines), while the dash line corresponds to a fiducial value $f_{gas} = 0.3$ with the associated scatter encompassing the range $f_{gas} = 0.2$–0.4 (i.e. the expected r.m.s. scatter, see Rodrigues et al. 2012; see grey dashed lines). We emphasize that these ranges of $Q_{gas}$ and $f_{gas}$ correspond to typical values expected in this range of mass and redshift (see Puech 2010; Puech et al. 2010).

As concluded by Johnson et al. (2018), it is difficult considering the full sample to discriminate between any of these models (or RDDs). Restricting the sample to VRDs (black points), it is interesting to note that the feedback-driven model can account for four of the five VRDs in the sample, while the gravity-driven model might be better suited to account for the remaining VRD. Given the small number of VRDs in the sample it remains difficult to draw any firm conclusion but we note that this tentative interpretation might be consistent with the suggestion of Krumholz, Kruijssen & Crocker (2017) of a possible transition between a gravity-driven regime at high $z$ and a feedback-driven regime at lower $z$ occurring at $z \sim 0.5$. While this might prevent to efficiently discriminate between the two models at the redshift of the IMAGES-CDFS sample ($z \sim 0.6$), it might also explain why the two sets of tracks considered *together* can reasonably account for the position of all VRDs in the $\sigma_{gas}$ versus *SFR* plane. In this interpretation, galaxies with high *SFR* values would have their $\sigma_{gas}$ mostly driven by feedback, while galaxies with smaller *SFR* ($\lesssim 10 M_{\odot}$ yr$^{-1}$) would have their $\sigma_{gas}$ mostly driven by gravity.[13] A larger sample of VRDs will be needed to confirm the above possible interpretation.

Since selecting high-$z$ VRDs requires a larger number of selection criteria, it results that the fraction of VRDs is expected to be smaller than the fraction of RDDs in a given galaxy population (see e.g. a quantified comparison at $z \sim 0.9$ in Rodrigues et al. 2017). The last panel of Fig. 7 reproduces the evolution of the fraction of RDDs within the population of star-forming galaxies (now considering a less strict selection with $V_{rot}/\sigma_{gas} > 1$) as fitted to a collection of studies from the literature by Tiley et al. (2016). This fraction determined in the present IMAGES (full) sample is consistent with this trend, which scales as $-0.2z$. We also show in this panel the evolution of the fraction of VRDs in the whole population of galaxies at different look-back times (i.e. further correcting for the fraction of star-forming galaxies at high-$z$ or the fraction of spirals amongst intermediate-mass galaxies at $z = 0$) as determined by H09 at $z \sim 0$ and $z \sim 0.6$ and Rodrigues et al. (2017) at $z \sim 0.9$. Interestingly, a faster decline is found, with $-0.5z$.

---

[12]We qualify them as 'virialized' because requiring both morphology and spatially resolved kinematics consistent with expectations from local spirals (see e.g. Rodrigues et al. 2017) allows separating truly isolated discs from, e.g. merger remnant that are in the process of reforming discs (see e.g. Hung et al. 2016).

[13]Note that this limit in *SFR* roughly translates into a depletion time-scale $\sim 1$ Gyr, which corresponds to a widely used limit for defining 'starbursts' (see Section 4.4).

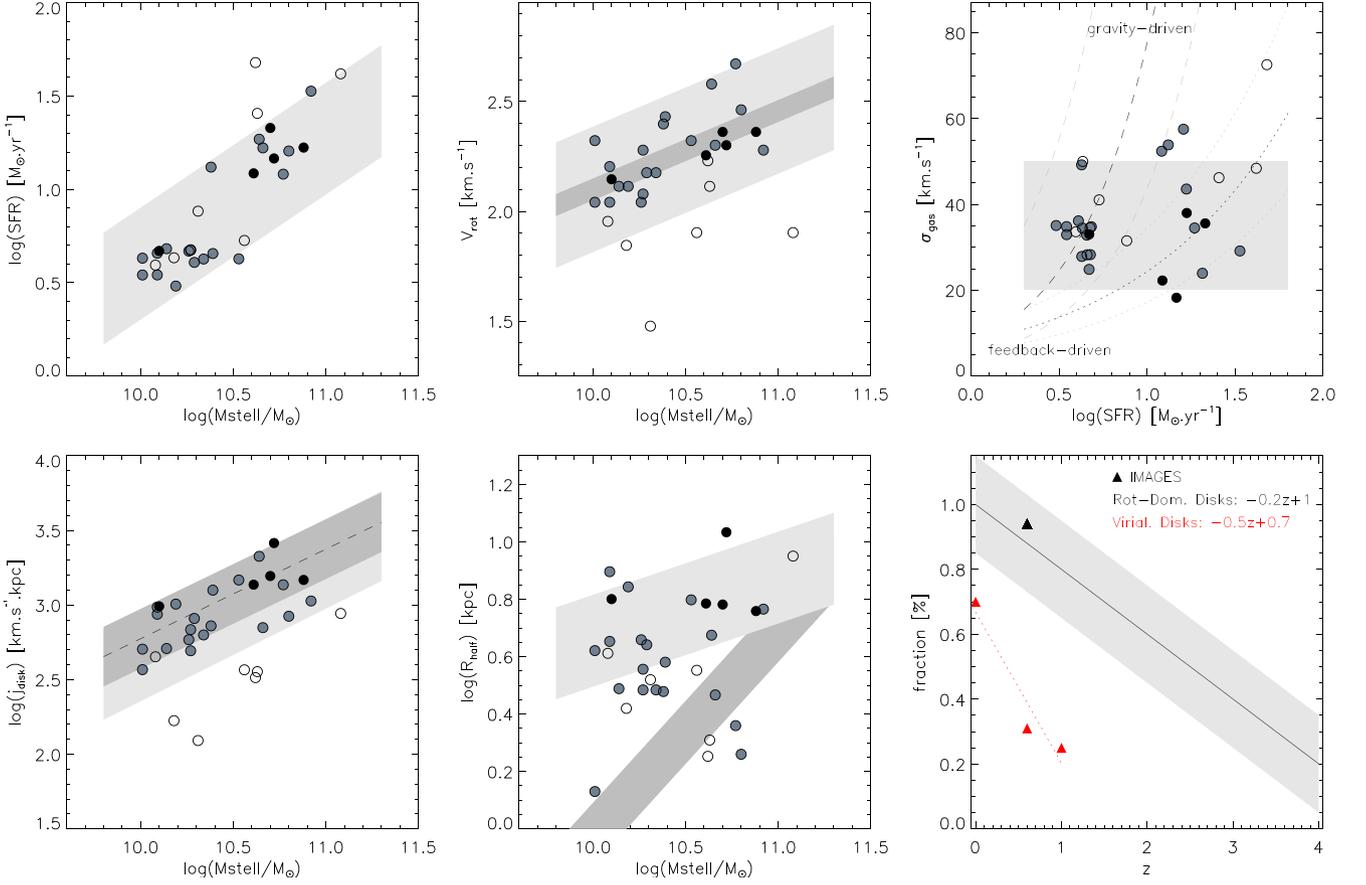

**Figure 7.** Main scaling relations in the full $z \sim 0.6$ IMAGES-CDFS sample. In all panels (except the last one), black symbols represent virialized rotating discs, grey symbols represent galaxies with $V_{\rm rot}/\sigma_{\rm gas} > 3$ (see Section 5.3), and open symbols represent all other galaxies (i.e. with significant morphological or kinematic perturbations). *From top to bottom and left to right:* SFR versus $M_{\rm stellar}$ (the grey area represents the locus of galaxies at similar redshift and mass using the fit of Dutton, van den Bosch & Dekel 2010), $V_{\rm rot}$ versus $M_{\rm stellar}$ (TFR; the dark grey area represents the fitted relation with associated residual scatter obtained by Puech et al. 2008 in the sample restricted to rotating discs, while the lighter grey area represents the scatter associated with the full sample), $\sigma_{\rm gas}$ versus SFR (the lighter grey area represents the scatter obtained in larger samples at similar redshift and mass, see Johnson et al. 2018; the dotted lines represents typical predictions for feedback-driven models, while the dashed lines represents typical predictions for gravity-driven models, see the text), $j_{\rm disc}$ versus $M_{\rm stellar}$ (Fall relation; the lighter grey area represents the typical scatter at similar mass and redshifts obtained in the full sample of Harrison et al. 2017, while the darker area and dash line, respectively, represent the local scatter and relation from Romanowsky & Fall 2012), and size versus $M_{\rm stellar}$ (the lighter and darker grey areas represent the locus of late and early-type galaxies at similar redshifts and mass from van der Wel et al. 2014; note that a significant number of late-type galaxies are also spread below the lighter grey area, which is not represented here for clarity). *Bottom rightmost panel:* evolution with $z$ of the fraction of rotation-dominated galaxies relative to the number of star-forming galaxies ($V_{\rm rot}/\sigma_{\rm gas} > 1$; the black line and the grey area represent the fitting to collection of results in the literature from Tiley et al. 2016, while the black triangle represents the fraction determined in the *full* IMAGES sample, see Section 3) versus fraction of virialized rotating discs relative to the overall population of galaxies of same mass (red symbols; the fractions at $z \sim 0$ and $z \sim 0.6$ were determined from H09, while the one at $z \sim 1$ was determined in Rodrigues et al. 2017, see the text).

## 6 CONCLUSIONS

Gas-rich major mergers are more and more recognized as an important driver for the structural evolution of galaxies at $z < 1$, including for the formation of ~50 per cent of present-day spiral galaxies. We studied in this paper the time-averaged evolution of the most important properties of galaxy along a typical $z \sim 0.6$ major mergers. The main results can be summarized as follows:

(i) The time variation of all the ISM, dynamical, and morphological properties inferred from observations provides a coherent description of the average major merger at $z \sim 0.6$ consistent with expectations from models. In particular, the *SFR*, gas surface density $\Sigma_{\rm gas}$, and velocity dispersion $\sigma_{\rm gas}$ are strongly correlated with co-phased enhancements. These are strongly correlated in time with important perturbations in the gravitational potential $\Delta\phi/\phi$;

(ii) These variations can be attributed to gas compressed by gravitational torques resulting of the interactions. Observations suggest that gravitational energy injected during the interaction can be converted into kinematic energy through shocks (reflected by enhanced electron densities);

(iii) On average, major mergers at $z \sim 0.6$ result in a loss of angular momentum (and dynamical support from rotation) during the most perturbed phases of the interaction (i.e. first encounter and fusion), which translates into compact morphologies. However, the resulting balance in specific angular momentum between the beginning and the end of the merger is a net increase by $\sim 0.1$ dex, consistent with predictions for wet major mergers in cosmological simulations;

(iv) The short total gas depletion time-scale during the most perturbed phases suggests that the internal gas reservoir is rapidly exhausted and subsequently replenished. The best weak evolution of the TFR reported by several studies implies that at least part of the newly accreted gas was gravitationally bound to stellar discs at $z \sim 0.6$–1.5;

(v) Using scaling relations predicted by numerical simulations, the predicted distribution in *B/T* for the merger remnants matches the fractions of early-type (E/S0) versus late-type (Sp) galaxies at $z = 0$ within only a few per cents. These fractions are also in good agreement with predictions from cosmological simulations (e.g. Font et al. 2017) and confirm that quiescent formation histories, conversely to what was long assumed, are not required to form disc galaxies similar to present-day spirals;

(vi) These theoretical predictions, associated with observations of both local and distant galaxies, now all provide a coherent picture in which $\sim 50$ per cent of present-day spiral galaxies with masses similar to the MW have reformed their discs following major mergers that occurred in the past 8–9 Gyr (see Puech et al. 2012). These recent mergers are expected to result in the formation of galaxies with a large diversity in morphologies, from early-type E/S0 to galaxies with morphologies as late as Sbc, depending on the exact mass ratios, gas fractions, and orbits;

(vii) Major mergers are found to significantly increase the scatter of several scaling relations such as the TFR between $V_{\rm rot}$ and $M_{\rm stellar}$, or the Fall relation between $j_{\rm disc}$ and $M_{\rm stellar}$. Selecting high-$z$ discs as rotation-dominated galaxies help to reduce this scatter but not entirely, while selecting them based both on the morphological and kinematical spatial distribution (VRDs, see Rodrigues et al. 2017) can remove all the measurement scatter resulting in scaling relations with scatters similar to that measured in local relations. The latter selection also appears to be more immune to possible systematic shifts that could remain hidden in the residual scatter when relying on the RDD criterion, which could affect, e.g. the evolution of the Fall relation.

## ACKNOWLEDGEMENTS

We thank A. Romeo for useful discussions. We also thank L. Athanassoula, R. Bacon, S. Charlot, F. Combes, T. Contini, and J.-G. Cuby for useful comments on a very preliminary version of some of the results presented in this paper.